\documentclass[prd,showpacs,,nofootinbib,showkeys,preprint]{revtex4}

\usepackage{amsmath}
\usepackage{amssymb}
\usepackage{graphicx}
\usepackage{rotating}
\usepackage{color} 

\newcommand{\AddrAHEP}{
  {\it AHEP Group, Instituto de F\'{\i}sica Corpuscular --
    C.S.I.C./Universitat de Val{\`e}ncia \\
    Edificio de Institutos de Paterna, Apartado 22085,
  E--46071 Val{\`e}ncia, Spain}}

\newcommand{\AddrWur}{%
Institut f\"ur Theoretische Physik und Astronomie, 
Universit\"at W\"urzburg\\
Am Hubland, 
97074 Wuerzburg}

\def\gsim{\raise0.3ex\hbox{$\;>$\kern-0.75em\raise-1.1ex\hbox{$\sim\;$}}}
\def\lsim{\raise0.3ex\hbox{$\;<$\kern-0.75em\raise-1.1ex\hbox{$\sim\;$}}}

\begin{document}

\begin{flushright}
IFIC/08-30\\
\end{flushright}

\title{Supersymmetric seesaw type II: CERN LHC and lepton flavour violating 
phenomenology}

\author{M.~Hirsch} \email{mahirsch@ific.uv.es}\affiliation{\AddrAHEP}
\author{W. Porod} \email{porod@physik.uni-wuerzburg.de}\affiliation{\AddrWur}
\author{S.~Kaneko} \email{satoru@ific.uv.es}\affiliation{\AddrAHEP}

\begin{abstract}
We study the supersymmetric version of the type-II seesaw mechanism 
assuming minimal supergravity boundary conditions. We calculate 
branching ratios for lepton flavour violating (LFV) 
scalar tau decays, potentially observable at 
the LHC, as well as LFV decays at low energy, such as $l_i \to l_j 
+ \gamma$ and compare their sensitivity to the unknown seesaw parameters. 
In the minimal case of only one triplet coupling to the standard model 
lepton doublets, ratios of LFV branching ratios can be related unambigously 
to neutrino oscillation parameters. We also discuss how measurements of 
soft SUSY breaking parameters at the LHC can be used to indirectly 
extract information of the seesaw scale. 
\end{abstract}


\maketitle

\section{Introduction}
\label{sec:int}

Neutrinos have mass and non-trivial mixing angles, as neutrino 
oscillation experiments have shown \cite{Fukuda:1998mi,Ahmad:2002jz,%
Eguchi:2002dm,Collaboration:2007zza,KamLAND2007}. If neutrinos are 
{\em Majorana} particles, their mass at low energy is described by 
a unique dimension-5 operator \cite{Weinberg:1979sa}
\begin{equation}\label{eq:dim5}
m_{\nu} = \frac{f}{\Lambda} (H L) (H L).
\end{equation}
Using only renormalizable interactions, there are only three tree-level 
realizations of this operator \cite{Ma:1998dn}. The first one is the 
exchange of a heavy fermionic singlet. This is the celebrated seesaw 
mechanism \cite{Minkowski:1977sc,seesaw,MohSen}, which we will call 
seesaw type-I. The second possibility is the exchange of a scalar 
triplet \cite{Schechter:1980gr,Cheng:1980qt}. This is commonly known 
as seesaw type-II. And lastly, one could also add one (or more) fermionic 
triplets to the field content of the SM \cite{Foot:1988aq}. This is 
called seesaw type-III in  \cite{Ma:1998dn}, although this nomenclature 
is not universally accepted \footnote{Barr and Dorsner \cite{Barr:2005ss}, 
for example, add additional singlets to the seesaw type-I. This version 
of the seesaw - which the authors call type-III - might be named 
``double seesaw, variant-II'' to distinguish it from the original 
double seesaw \cite{Mohapatra:1986bd}, see also the related work 
in \cite{Akhmedov:1995ip}.}.

The dimension-5 operator of eq. (\ref{eq:dim5}) could also be generated 
at loop level. As the classical examples for loop generated neutrino 
masses we only mention the Zee model \cite{Zee:1980ai} (1-loop) 
and the Babu-Zee model \cite{Zee:1985id} (2-loop), although many more 
models exist in the literature. A list of generic 1-loop realizations 
of  eq. (\ref{eq:dim5}) can also be found in \cite{Ma:1998dn}.

At ``low'' energies one can neither decide whether tree-level or 
loop physics generates eq. (\ref{eq:dim5}), nor can any measurements 
of neutrino angles, phases or masses distinguish between the 
different tree-level realizations of the seesaw discussed above. 
Observables outside the neutrino sector are needed to ultimately 
learn about the origin of eq. (\ref{eq:dim5}). For loop generated 
neutrino masses, $f$ in eq. (\ref{eq:dim5}) can be a very small 
number and the scale $\Lambda$ at which new physics appears can be 
quite low, probably accessible at future accelerators such as the 
LHC or an ILC. The ``classical'' tree-level realizations of the seesaw, 
unfortunately, can not be put to the test in such a direct way. 
This can be straightforwardly understood by inverting eq. (\ref{eq:dim5}), 
which results in $\Lambda \sim f \Big(\frac{\rm 0.05 \hskip1mm eV}
{m_{\nu}}\Big) 10^{15}$ GeV. 

Indirect inside into the high-energy world might be possible in 
supersymmetric versions of the seesaw. In the renormalization group 
equations for the soft SUSY breaking slepton mass parameters terms 
proportional to the neutrino Yukawa couplings appear. 
If the scale where the right-handed neutrinos and/or the triplet 
decouple is below the scale at which SUSY breaks, lepton flavour 
violating (LFV) entries in the Yukawa matrices then induce LFV 
off-diagonals in the slepton mass matrices. This effect potentially 
leads to large values for lepton flavour violating lepton decays, 
such as $\mu \to e+\gamma$, even if the soft masses are completely 
flavour blind at high scale, as was first pointed out for the case 
of seesaw type-I in \cite{Borzumati:1986qx}. It is maybe not surprising 
then that with the increasingly convincing experimental evidence for 
non-zero neutrino masses a number of articles have studied the 
prospects for observing LFV processes, both at low energies and at 
future colliders, within the supersymmetric seesaw 
\cite{Hisano:1995nq,Hisano:1995cp,Ellis:2002fe,Deppisch:2002vz,%
Arganda:2005ji,Antusch:2006vw,Arganda:2007jw,Hisano:1998wn,%
Blair:2002pg,Freitas:2005et,Petcov:2003zb,Pascoli:2003rq,Petcov:2005yh,%
Petcov:2006pc}.

Despite the fact that a minimal seesaw type-II has fewer free 
parameters than the seesaw type-I, type-I seesaw has received 
considerably more attention in the literature. Probably one of 
the reasons for this preference is gauge coupling unification. 
As is well known \cite{Amaldi:1991zx,Ellis:1991ri},
 the SM gauge couplings unify within 
the minimal supersymmetric standard model (MSSM) at a scale 
around $M_G \simeq 2 \times 10^{16}$ GeV, if the SUSY particles 
have masses around the electro-weak scale. Adding gauge singlets 
does not destroy this nice feature of the MSSM. However, a scalar 
triplet with mass below the GUT scale changes the running of $g_1$ 
and $g_2$ in an unwanted way and gauge coupling unification is lost 
\cite{Rossi:2002zb}. A simple way to cure 
this defect of the seesaw-II consists in adding only complete $SU(5)$
multiplets (or GUT multiplets which can be decomposed into complete $SU(5)$
multiplets) to the standard model particle content. In this way the 
scale where couplings unify remains the same (at one loop level),
only the value of the GUT coupling changes \cite{Langacker:1980js}.

In this paper we calculate lepton flavour violating branching ratios 
of the scalar tau as well as LFV lepton decays at low energies, 
such as $l_i \to l_j + \gamma$ and $l_i \to 3 l_j$. For definiteness, 
we assume minimal Supergravity (mSugra) boundary conditions and fit 
the observed neutrino masses by a seesaw mechanism of type-II. We 
will discuss two different realizations. The first one is based 
on adding one pair of triplets to the MSSM, from which only one couples 
to the standard model leptons. This is the simplest supersymmetric 
version of the type-II seesaw. The second model we consider consists 
in adding a pair of ${\bf 15}$ and $\bf\overline{15}$ multiplets to 
the MSSM particle content \cite{Rossi:2002zb}. This second option 
allows to maintain gauge coupling unification also for $M_{15} \ll 
M_G$. 

We compare the sensitivities of low-energy and accelerator measurements 
and study their dependence on the unknown seesaw and SUSY parameters. 
Absolute values of LFV stau decays and LFV lepton decays depend very 
differently on the unknown SUSY parameters. For a light SUSY spectrum, 
say slepton masses below 200 GeV, the current upper bound on 
Br($\mu\to e+ \gamma$) limits seriously the possibility to observe LFV 
scalar tau decays. However, for heavier sparticles low energy data 
very rapidly looses its constraining power and large LFV at the LHC 
is allowed by current data. 

While absolute values of LFV observables depend very strongly on the 
soft SUSY breaking parameters, we discuss how ratios of LFV branching 
ratios can be used to eliminate most of the dependence on the unknown 
SUSY spectrum. I.e.~ratios such as, for example, Br(${\tilde\tau}_2\to 
e + \chi^0_1$)/Br(${\tilde\tau}_2\to \mu + \chi^0_1$) are constants for 
fixed neutrino parameters over large parts of the supersymmetric parameter 
space. Measurements of such ratios would allow to extract valuable 
information about the seesaw parameters: In the minimal type-II seesaw 
case these ratios can be calculated as function of measurable low-energy 
neutrino data. For the more involved case of the ${\bf 15}+\overline{\bf 15}$ 
model this simple connection is lost in general, but relations to 
neutrino data can be (re-) established in some simple, extreme cases 
for the Yukawa matrix ${\bf Y}_{15}$. We therefore study such ratios in 
some detail, first analytically then numerically. 

The presence of new non-singlet states below the GUT scale does not only 
affect the running of gauge couplings but also the evolution of the soft 
SUSY breaking parameters. Measurements of soft SUSY masses at the LHC 
and at a possible ILC therefore contain indirect information about the 
physics at higher energy scales \cite{Blair:2002pg,Deppisch:2007xu}. 
>From the different soft scalar and gaugino masses one can define certain 
``invariants'', 
i.e. parameter combinations which are nearly constant over large ranges 
of the mSugra parameter space \cite{Buckley:2006nv}, at least in leading 
order approximation. If the measured values of all the  invariants depart 
from the mSugra expectation in a consistent way, one could gain some 
indirect estimate of the mass scale of the new particles, the scale of 
the seesaw type-II. We discuss first some leading order analytical 
approximation, before showing by numerical calculation the limitations 
of the simplified analytical approach. While the different invariants 
indeed contain useful information about the high energy physics, reliable 
quantitative conclusions about the mass scale of the ${\bf 15}$ require 
highly precise measurements of soft masses as well as a full numerical 
2-loop analysis.

The rest of this paper is organized as follows. In the next section 
we will recall the basic features of the supersymmetric seesaw type-II 
and discuss a $SU(5)$ motivated variant, which adds a pair of ${\bf 15}$ 
and $\overline{\bf 15}$. Section \ref{sec:ana} then discusses analytical 
solutions for the RGEs and presents estimates for slepton mixing angles 
and the corresponding LFV observables. In Section \ref{sec:num} we present 
our numerical results for LFV decays at low energies and accelerators. This 
numerical study demonstrates the reliability of our analytical approximations 
for the LFV observables. We then discuss soft masses and the seesaw type-II 
scale, demonstrating by a numerically exact calculation that for soft 
masses the leading order approximations are not accurate enough to draw 
quantitative conclusions. We then summarize in section \ref{sec:cncl}.

\section{Setup: mSugra with seesaw type II}
\label{sec:setup}

In this section, to set up the notation, we briefly recall the main 
features of the seesaw type-II and mSugra. We then outline a simple 
$SU(5)$ motivated model based on the work of \cite{Rossi:2002zb}. 

\subsection{Supersymmetric seesaw with triplet(s)}
\label{sec:triplet}

In supersymmetry at least two $SU(2)$ triplet states $T_{1,2}$ with 
opposite hypercharge are needed to cancel anomalies. Thus, the minimal 
SUSY potential including triplets can be written as 
\begin{equation}\label{triplet_pot}
W = W_{\rm MSSM} + \frac{1}{\sqrt{2}}\Big(Y_T^{ij}L_i T_1 L_j + 
\lambda_1 H_1 T_1 H_1 + \lambda_2 H_2 T_2 H_2\Big) + M_T T_1 T_2. 
\end{equation}
Here $T_1$ ($T_2$) are supermultiplets with hypercharge $Y=1$ 
($Y=-1$) and $H_{1,2}$ are the standard Higgs doublets with 
$Y=\mp 1/2$. The matrix $Y_T$ is complex symmetric, $\lambda_{1,2}$ 
are arbitrary constants and $M_T$ gives mass to the triplets, 
supposedly at a very high scale. Note that only $T_1$ couples 
to the SM leptons, thus in the minimal (supersymmetric) model with 
two triplets the only source of lepton flavour violation 
resides in the matrix $Y_T$. 

Integrating out the heavy triplets at their mass scale the dimension-5 
operator of eq.~(\ref{eq:dim5}) is generated and after electro-weak 
symmetry breaking the resulting neutrino mass matrix can be written as 
\begin{eqnarray}\label{eq:ssII}
m_\nu=\frac{v_2^2}{2} \frac{\lambda_2}{M_T}Y_T.
\end{eqnarray}
where $v_2$ is the vacuum expectation value of Higgs doublet $H_2$ and 
we use the convention $\langle H_i\rangle = \frac{v_i}{\sqrt{2}}$. 
Note that eq. (\ref{eq:ssII}) depends on the energy scale. $m_\nu$ 
is measured at low energies, whereas for the calculation of $m_\nu$ we 
need to know $\lambda_2$, $Y_T$ and $M_T$ as input paramters at the high 
scale. One can use an iterative procedure to find the high scale parameters 
from the low energy measured quantities, as explained in section 
\ref{sec:num}. In the basis where the charged lepton masses are diagonal, 
eq. (\ref{eq:ssII}) is diagonalized by 
\begin{eqnarray}
{\hat m}_\nu = U^T \cdot m_{\nu} \cdot U,
\end{eqnarray}
where the neutrino mixing matrix $U$ is, in standard notation \cite{pdg}, 
given by
\begin{eqnarray}\label{def:unu}
U=
\left(
\begin{array}{ccc}
 c_{12}c_{13} & s_{12}c_{13}  & s_{13}e^{-i\delta}  \\
-s_{12}c_{23}-c_{12}s_{23}s_{13}e^{i\delta}  & 
c_{12}c_{23}-s_{12}s_{23}s_{13}e^{i\delta}  & s_{23}c_{13}  \\
s_{12}s_{23}-c_{12}c_{23}s_{13}e^{i\delta}  & 
-c_{12}s_{23}-s_{12}c_{23}s_{13}e^{i\delta}  & c_{23}c_{13}  
\end{array}
\right) \times
\left(
\begin{array}{ccc}
e^{i\alpha_1/2} & 0 & 0 \\
0 & e^{i\alpha_2/2}  & 0 \\
0 & 0 & 1
\end{array}
\right) .
\end{eqnarray}
Here $s_{ij}\equiv \sin\theta_{ij}$ ($c_{ij}=\cos\theta_{ij}$). For 
Majorana neutrinos, $U$ contains three phases: $\delta$ is 
the (Dirac-) CP violating phase, which appears in neutrino 
oscillations, and $\alpha_{1,2}$ are Majorana phases, which 
can only be observed in lepton number violating processes. 
Neutrino oscillation experiments can be fitted with either 
a normal hierarchical spectrum (NH), or with inverted hierarchy 
(IH). If one does not insist in ordering the neutrino mass eigenstates 
$m_{\nu_i}$, $i=1,2,3$ with respect to increasing mass, the matrix 
$U$ can describe both possibilities without re-ordering of angles. 
In this convention, which we will use in the following, $m_{\nu_1} 
\simeq 0$ ($m_{\nu_3}\simeq 0$) corresponds to normal (inverse) 
hierarchy and $s_{12}$, $s_{13}$ and $s_{23}$ are the solar 
($s_{\odot}$), reactor ($s_{R}$) and atmosperic angle ($s_{\rm Atm}$) 
for both type of spectra.

Note that 
\begin{equation}\label{diagYT}
{\hat Y}_T = U^T \cdot Y_T \cdot U
\end{equation}
i.e. $Y_T$ is diagonalized by {\em the same matrix as $m_{\nu}$}. 
If all neutrino eigenvalues, angles and phases were known, $Y_T$ 
would be fixed up to an overall constant which can be easily 
estimated to be 
\begin{equation}\label{est}
\frac{M_T}{\lambda_2} \simeq 10^{15} {\rm GeV} \hskip2mm 
\Big(\frac{0.05 \hskip1mm {\rm eV}}{m_{\nu}}\Big).
\end{equation}
At this points it might be worth recalling the main differences between 
seesaw type-II and seesaw type-I. In seesaw type-I there is one 
non-zero mass eigenstate for the light neutrinos for each right-handed 
neutrino added to the model. In contrast, seesaw-II can produce three 
non-zero neutrino masses with only one triplet. Thus the minimal model 
for seesaw type-II with only one triplet coupling to $L$ has less 
parameters than seesaw type-I. We can count the new parameters in 
eq. (\ref{triplet_pot}): $Y_T$ being complex symmetric has 9 parameters. 
Additionally we have $\lambda_{1,2}$ and $M_T$. All three could 
in principle be complex. However, field redefinitions on $T_1$ and 
$T_2$ can be applied to remove two of the three phases, thus there 
is a a total of 13 parameters. Note, however, that only 11 of them are 
related to neutrino physics. Since we have the freedom to write down 
eq. (\ref{triplet_pot}) 
in the basis, where the charged lepton mass matrix is diagonal, we 
only have to add three charged lepton masses to the counting of free 
parameters.\footnote{In the non-supersymmetric version of seesaw-II  
$\frac{\lambda_2}{M_T}\rightarrow \frac{\mu}{M_T^2}$, with $\mu$ having 
dimension of mass, but the number of parameters related with neutrino 
physics does not change.} 
This number should be compared to the 21 free parameters in seesaw 
type-I for three right-handed neutrinos \cite{Santamaria:1993ah}. 
At low energies a maximum of 12 parameters can be fixed by measuring 
lepton properties: 3 neutrino and 3 charged lepton masses, 3 angles and 
3 phases. Thus from neutrino data neither seesaw type-II nor seesaw 
type-I can be completely reconstructed. However, especially important 
in the following is the fact, see eq. (\ref{diagYT}), that low-energy 
neutrino angles are directly related to the high-energy Yukawa matrix 
in seesaw-II, whereas no such simple connection exists in the seesaw 
type-I, see also the discussion in \cite{Ellis:2002fe}.

\subsection{$SU(5)$ inspired model with {$\bf 15$}+{$\overline{\bf 15}$}}
\label{sec:su5}

In this section we outline the basics of an $SU(5)$ inspired 
model, which adds a pair of ${\bf 15}$ and $\overline{\bf 15}$ 
to the MSSM particle spectrum \cite{Rossi:2002zb}. Our numerical 
calculations will all be based on this variant, since it allows 
to maintain gauge coupling unification for $M_T\ll M_G$, as 
discussed in the introduction. 

Under $SU(3)\times SU_L(2) \times U(1)_Y$ the ${\bf 15}$ 
decomposes as 
\begin{eqnarray}\label{eq:15}
{\bf 15} & = &  S + T + Z \\ \nonumber
S & \sim  & (6,1,-\frac{2}{3}), \hskip10mm
T \sim (1,3,1), \hskip10mm
Z \sim (3,2,\frac{1}{6}).
\end{eqnarray}
$T$ has the same quantum numbers as the triplet $T_1$ discussed above. 
The $SU(5)$ invariant superpotential reads as 
\begin{eqnarray}\label{eq:pot15}
W & = & \frac{1}{\sqrt{2}}{\bf Y}_{15} {\bar 5} \cdot 15 \cdot {\bar 5} 
   + \frac{1}{\sqrt{2}}\lambda_1 {\bar 5}_H \cdot 15 \cdot {\bar 5}_H 
+ \frac{1}{\sqrt{2}}\lambda_2 5_H \cdot \overline{15} \cdot 5_H 
+ {\bf Y}_5 10 \cdot {\bar 5} \cdot {\bar 5}_H \\ \nonumber
 & + & {\bf Y}_{10} 10 \cdot 10 \cdot 5_H + M_{15} 15 \cdot \overline{15} 
+ M_5 {\bar 5}_H \cdot 5_H
\end{eqnarray}
Here, ${\bar 5}=(d^c,L)$, $10=(u^c,e^c,Q)$, ${5}_H =(t,H_2)$ and 
${\bar 5}_H=({\bar t},H_1)$. Below the GUT scale in the $SU(5)$-broken 
phase the potential contains the terms 
\begin{eqnarray}\label{eq:broken}
 &  & \frac{1}{\sqrt{2}}(Y_T L T_1 L +  Y_S d^c S d^c) 
+ Y_Z d^c Z L + Y_d d^c Q H_1 + Y_u u^c Q H_2  
+ Y_e e^c L H_1 \\ \nonumber
& + & \frac{1}{\sqrt{2}}(\lambda_1 H_1 T_1 H_1 +\lambda_2 H_2 T_2 H_2) 
+ M_T T_1 T_2 + M_Z Z_1 Z_2 + M_S S_1 S_2 + \mu H_1 H_2 
\end{eqnarray}
The first term in eq. (\ref{eq:broken}) is responsible for the generation 
of the neutrino masses in the same way as discussed for the triplet-only 
case in the previous subsection. $Y_d$, $Y_u$ and $Y_e$ 
generate quark and charged lepton masses in the usual manner.  However, 
in adddition there are the matrices $Y_S$ and $Y_Z$, which, in  
principle,  are not determined by any low-energy data. In the calculation 
of LFV observables in supersymmetry both matrices, $Y_T$ and 
$Y_Z$, contribute. For the case of a complete ${\bf 15}$, apart 
from threshold corrections, $Y_T=Y_S=Y_Z$. One can recover the 
results for the simplest triplet-only model, as far as lepton 
flavour violation is concerned, by putting $Y_S=Y_Z=0$.

As long as $M_Z \sim M_S \sim M_T \sim M_{15}$ gauge coupling unification 
will be mantained. The equality need not be exact for successful unification. 
In our numerical studies we have taken into account the different running
of these mass parameters but we decouple them all at the scale $M_T(M_T)$
because the differences are small.

\section{Analytical results}
\label{sec:ana}

\subsection{Approximate solutions for the RGEs}

In mSugra one has in total five parameters at the GUT scale 
\cite{Haber:1984rc}. These are usually chosen to be $M_0$, the 
common scalar mass, $M_{1/2}$, the gaugino mass parameter, $A_0$, 
the common trilinear parameter, 
$\tan\beta=\frac{v_2}{v_1}$ and the sign of $\mu$. 
For the full set of RGEs for the ${\bf 15}$ + $\overline{\bf 15}$ see 
\cite{Rossi:2002zb}. In the numerical calculation, presented in 
the next section, we solve the exact RGEs. However, the following 
approximative solutions are very helpful in gaining a qualitative 
understanding. 

The gauge couplings are given as 
\begin{eqnarray}
\alpha_1(m_Z) &=& \frac{5 \alpha_{em}(m_Z)}{3 \cos^2\theta_W}, \hspace{1cm}
\alpha_2(m_Z) = \frac{\alpha_{em}(m_Z)}{\sin^2\theta_W}, \\  \nonumber 
\alpha_i(m_{SUSY}) &=& \frac{\alpha_i(m_Z)}{1- \frac{\alpha_i(m_Z)}
               {4 \pi} b_i^{SM} \log{\frac{m_{SUSY}^2}{m_Z^2}}}, \\ \nonumber
\alpha_i(M_T) &=& \frac{\alpha_i(m_{SUSY})}{1- \frac{\alpha_i(m_{SUSY})}
               {4 \pi} b_i \log{\frac{M_T^2}{m_{SUSY}^2}}}, \\ \nonumber
\alpha_i(M_G) &=& \frac{\alpha_i(M_T)}{1- \frac{\alpha_i(M_T)}
               {4 \pi} (b_i+\Delta b_i) \log{\frac{M_G^2}{M_T^2}}}.
\end{eqnarray}
$b^{SM}=(b_1,b_2,b_3)^{SM} = \textstyle (\frac{41}{10},-\frac{19}{6},-7)$  
for SM and $b=(b_1,b_2,b_3)^{MSSM}= \textstyle (\frac{33}{5},1,-3)$  
for MSSM. $M_T$ denotes the mass of the triplet (15-plet). For the case 
of the complete 15-plet one finds $\Delta b_i=7$ whereas for the case 
with triplets-only one finds $\Delta b_1=18/5$, $\Delta b_2=4$ and 
 $\Delta b_3=0$.
Using the equality $\alpha_1(M_G) = \alpha_2(M_G)$ determines the 
GUT-scale $M_G$ via
\begin{eqnarray}
\log \frac{M_G^2}{M_T^2} &=&
\frac{1}{\alpha_1(m_{SUSY}) \alpha_2(m_{SUSY}) 
        (b_1 + \Delta b_1 - b_2 - \Delta b_2 )} \\ \nonumber 
&& \cdot \left( 4 \pi (\alpha_2(m_{SUSY}) - \alpha_1(m_{SUSY}))
 + \alpha_1(m_{SUSY}) \alpha_2(m_{SUSY}) (b_2- b_1 ) 
\log{\frac{M_T^2}{m_{SUSY}^2}} \right)  
\end{eqnarray}
Note, that in the case of the complete 15-plet $M_G$ is independent 
of $M_T$. For the gaugino masses one finds 
\begin{eqnarray}
M_i(m_{SUSY}) = \frac{\alpha_i(m_{SUSY})}{\alpha(M_G)} M_{1/2}.
\label{eq:gaugino}
\end{eqnarray}
Eq. (\ref{eq:gaugino}) implies that the ratio $M_2/M_1$, which is 
measured at low-energies, has the usual mSugra value, but the 
relationship to $M_{1/2}$ is changed.
Neglecting the Yukawa couplings ${\bf Y}_{15}$, see below, for the 
soft mass parameters of the first two generations one obtains
\begin{eqnarray}\label{eq:scalar}
m_{\tilde f}^2  &=& M_0^2  +  
\sum_{i=1}^3  c^{\tilde f}_i \left(
\left(\frac{\alpha_i(M_T)}{\alpha(M_G)}\right)^2 f_i
 + f_i'
\right) M_{1/2}^2, \\ \nonumber
 f_i &=& \frac{1}{b_i} \left(
1 - {\left[1 + \frac{\alpha_i(M_T)}{4 \pi} b_i \log
\frac{M^2_T}{m_Z^2}\right]^{-2}} \right), \\
   f_i' &=& \frac{1}{b_i+\Delta b_i} \left(
1 - {\left[1 + \frac{\alpha(M_G)}{4 \pi} (b_i + \Delta b_i) \log
\frac{M_G^2}{M_T^2}\right]^{-2}} \right).
\end{eqnarray}
The various coefficients $c^{\tilde f}_i$ are given 
in table~\ref{tab:coeff}.
\begin{table}[ht]
\begin{tabular}{|c|ccccc|}
\hline
$\tilde f$ &   $\tilde E$ & $\tilde L$ &$\tilde D$ & $\tilde U$ & $\tilde Q$ \\
\hline
$c^{\tilde f}_1$ & $\frac{6}{5}$ & $\frac{3}{10}$ & $\frac{2}{15}$
                 & $\frac{8}{15}$ & $\frac{1}{30}$ \\
$c^{\tilde f}_2$ & 0 &  $\frac{3}{2}$ &  0 & 0 & $\frac{3}{2}$ \\
$c^{\tilde f}_3$ & 0 & 0 &  $\frac{8}{3}$ & $\frac{8}{3}$ & $\frac{8}{3}$ 
\\ \hline
\end{tabular}
\caption{Coefficients $c^{\tilde f}_i$  for 
eq.~(\ref{eq:scalar}).}
\label{tab:coeff}
\end{table}

Individual SUSY masses depend strongly on the initial values for 
$M_0$ and $M_{1/2}$. However, one can form different combinations, 
such as
\begin{eqnarray}\label{eq:exaInv}\nonumber
(m_{\tilde L}^2 -m_{\tilde E}^2)/M_1^2 & = & 
\left(\frac{\alpha({M_G})}{\alpha_1(m_{SUSY})}\right)^2
\Big(\frac{3}{2}\left[\left(\frac{\alpha_2(m_T)}{\alpha(m_G)}\right)^2 
               f_2+f_2'\right] 
- \frac{9}{10}\left[\left(\frac{\alpha_1(m_T)}{\alpha(m_G)}\right)^2
    f_1+f_1'\right] \Big), 
\end{eqnarray}
which, to first approximation, are constants over large regions of 
mSugra space. We will call such combinations ``invariants''.

\begin{figure}[t] \centering
\includegraphics[height=60mm,width=80mm]{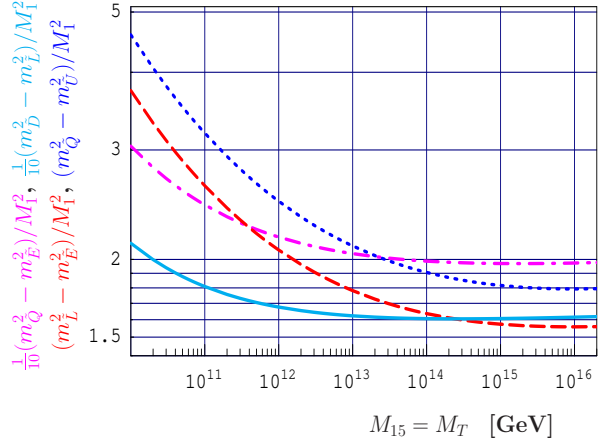} 

\vskip0mm
\caption{Four different ``invariant'' combinations of soft masses (left) 
versus the mass of the ${\bf 15}$-plet, $M_{ 15}=M_T$. The plot 
assumes that the Yukawa couplings ${\bf Y}_{15}$ are negligibly small. 
The calculation is at 1-loop order in the leading-log approximation. 
}
\label{fig:ana}
\end{figure}

Figure (\ref{fig:ana}) shows four different invariants 
as a function of $M_{15}=M_T$, calculated using eqs (\ref{eq:gaugino}) 
- (\ref{eq:scalar}). For $M_T=M_G$ one reaches the mSugra limit. For 
lower values of $M_T$ one obtains a logarithmic dependence on the 
value of $M_T$. If all the different invariants depart from their 
mSugra values in a consistent way, measurements of these parameter 
combinations can be used to obtain indirect information about the 
seesaw scale.
 In practice the ``invariants'' do depend on the 
SUSY spectrum and thus, indirectly still depend to some degree on 
the initial values of $M_0$ and $M_{1/2}$. We will discuss this point 
in more details in the numerical section.


For the off-diagonal elements of the slepton mass matrix, we will 
discuss only the left sector, since right slepton mass parameters 
do not run to first order approximation \cite{Rossi:2002zb}. In our 
numerical calculation we do solve the RGEs exactly and confirm this 
expectation. 
Off-diagonal elements are induced in $m^2_{\tilde L}$ due to the non-trivial 
flavour structure of the matrices $Y_T$ and $Y_Z$. $Y_T$ and $Y_Z$ 
appear symmetrically in the RGEs \cite{Rossi:2002zb}. Since only 
$Y_T$ can be fixed from low-energy data, for a general $Y_Z$ the 
off-diagonal entries of $m^2_{\tilde L}$ do not follow any correlation with 
low-energy physics. For this reason in the following we will consider 
two extreme cases: (a) $Y_Z=Y_T$, we will call this the 15-plet case; 
and (b) $Y_Z=0$, we will refer to this as the triplet case.

For $m^2_{\tilde L}$ one finds the following 
approximation in the case of the 15-plet:
\begin{eqnarray}\label{mslep15}
\Delta m^2_{{\tilde L},ij} &=& 
- \frac{1}{16\pi^2}  \left(Y^\dagger_T Y_T\right)_{ij}
\int^{\log \frac{M_G^2}{M_T^2}}_0
\left(
18 M^2_0 +\left(  \frac{34}{5} f'_1(t)  + 30 f'_2(t) + 16 f'_3(t)\right)  
M^2_{1/2} \right. \nonumber  \\
&& \hspace{4cm}  + 
3 (A_0 - \frac{9}{68} M'_1(t) - \frac{7}{8} M'_2(t))^2 \nonumber  \\
&& \hspace{4cm} \left.  
+ 3 (A_0 - \frac{7}{204} M'_1(t) - \frac{3}{8} M'_2(t) - \frac{4}{3} M'_3(t))^2
\right) dt \\
M'_i(t) &=& M_{1/2}\left( 1- \frac{1}{1 + \frac{1}{4 \pi} 
(b_i + \Delta b_i) \alpha(M_G) t} \right)
\end{eqnarray}
In case of the triplet one finds
\begin{eqnarray}\label{mslepT}
\Delta m^2_{{\tilde L},ij} &=& 
- \frac{1}{16\pi^2}  \left(Y^\dagger_T Y_T\right)_{ij}
\int^{\log \frac{M_G^2}{M_T^2}}_0
\left(9 M^2_0 + \left( \frac{27}{5} f'_1(t) + 21 f'_2(t) \right)  
M^2_{1/2} \right. \nonumber  \\
&& \hspace{4cm}  + \left.
3 (A_0 - \frac{9}{68} M'_1(t) - \frac{7}{8} M'_2(t))^2 
\right) dt 
\end{eqnarray}
The integration over $t$ can be done analytically leading to
corrections to the formulas for $(\Delta m^2_{\widetilde{L}})_{ij}$. 

We have found that the approximation formulas shown above 
work less well than the corresponding formulas for the seesaw type-I 
case and only give a rough order of magnitude estimate. The reason for 
this difference is, that in seesaw type-I the $Y_\nu$ hardly run, unless 
left neutrinos are very degenerate, as either the Yukawas themselfes 
are small or in the case of large Yukwas the contribution from gauge 
and (top) Yukawa couplings nearly cancel each other. Such a cancellation 
does not take place in case of $Y_T$, thus leading to significantly 
stronger dependence of $Y_T$ on the renormalization scale and consequently 
larger differences between the numerical solutions and eqs (\ref{mslep15}) 
and (\ref{mslepT}).

However, we have found that it is possible to improve the accuracy 
of the approximation formulas using the results of the next subsection, 
see eq. (\ref{expressYuk}). The idea here is to replace the running 
Yukawa coupling $Y_T$ by the measured low-energy neutrino masses and 
angles times the unkonw coupling $\lambda_2$. In case $\lambda_2$ is 
sufficiently small, this parameter does run very little and eqs 
(\ref{mslep15}) and (\ref{mslepT}) agree already very well with 
the numerical results. For large values of $\lambda_2$, we can find 
approximate solutions for the RGE for this parameter, following 
the procedure outlined in \cite{Blair:2002pg}. 

We define
\begin{eqnarray}
X\equiv
\frac{\lambda_2^2}{4\pi},\ \ 
Y_t\equiv
\frac{y_t^2}{4\pi}.
\end{eqnarray}
The solution for the RGE for $\lambda_2$ is then given in terms of X(t) 
by ($t=\log\frac{M_G^2}{Q^2}$)
\begin{eqnarray}\label{solX}
X(t)&=&\frac{X(M_G)u_{X}(t)}{1+\frac{7}{2\pi}X(M_G)\int^t_0 u_X(t')dt' }
,\\ \nonumber
u_X(t')&=&\frac{
(1+\frac{6}{2\pi}Y_t(M_G)t')^{1/\frac{6}{2\pi}}
}{
E_X(t')}
,\\ \nonumber
E_X(t')&=&
\left(1+\frac{b_1+\Delta b_1}{2\pi} 
\alpha_1(M_G)t' \right)^{\frac{1}{2\pi}\frac{9}{5}/(b_1+\Delta b_1)}
\left(1+\frac{b_2+\Delta b_2}{2\pi} \alpha_2(M_G)t' \right)^{\frac{1}{2\pi}7/
(b_2+\Delta b_2)}.
\end{eqnarray}
we have found that, assuming an approximately constant $Y_t$, the 
above equations become easy to solve and describe the running of 
$\lambda_2$ to a rather good approximation. Eq. (\ref{solX}) and 
eq. (\ref{expressYuk}), together with eqs (\ref{mslep15}) and 
(\ref{mslepT}) then allow to estimate LFV entries in $m^2_{\tilde L}$ 
up to an accuracy of typically a few percent.

\subsection{Analytical results for flavour violating processes}

Here we concentrate exclusively on the left-slepton sector. Taking into 
account the discussion given above this is expected to be a reasonable 
first approximation. The left-slepton mass matrix is diagonalized by a 
matrix $R^{\tilde l}$, which in general can be written as a product of 
three Euler rotations.  However, if the mixing between the different 
flavour eigenstates is sufficiently small, the three differenct angles 
can be estimated by the following simple formula
\begin{equation}\label{slepAng}
\theta_{ij} \simeq \frac{(\Delta m_{\tilde L}^2)_{ij}}
       {(\Delta m_{\tilde L}^2)_{ii} - (\Delta m_{\tilde L}^2)_{jj}}.
\end{equation}
LFV decays are directly proportional to the squares of these mixing 
angles as long as all angles are small. Taking the ratio of two decays, 
for example, 
\begin{equation}\label{mainresult}
\frac{Br({\tilde\tau}_2 \to  e +\chi^0_1)}
     {Br({\tilde\tau}_2 \to  \mu +\chi^0_1)}
 \simeq \Big(\frac{\theta_{{\tilde e}{\tilde\tau}}}
                   {\theta_{{\tilde\mu}{\tilde\tau}}}\Big)^2 
 \simeq 
\Big(\frac{(\Delta m_{\tilde L}^2)_{13}}{(\Delta m_{\tilde L}^2)_{23}}\Big)^2,
\end{equation}
one expects that (a) all the unknown SUSY mass parameters and (b)
the denominators of eq.~(\ref{slepAng}) cancel approximately. To calculate 
estimates for different ratios of branching ratios we define
\begin{equation}
\label{eq:def-rijkkl}
r^{ij}_{kl}\equiv \frac{|(\Delta m_{\tilde{L}}^2)_{ij}|}
                      {|(\Delta m_{\tilde{L}}^2)_{kl}|}
\end{equation}
where the observable quantities are $(r^{ij}_{kl})^2$. Of course, only
two of the three possible combinations that can be formed are
independent. 

We next derive some analytical formulas for $(r^{ij}_{kl})^2$ in 
terms of observable neutrino parameters. 
The neutrino Yukawa coupling $Y_T$ can be written in terms of 
observable parameters 
\begin{eqnarray}
Y_T=\frac{2 M_T}{v_2^2 \lambda_2}m_\nu
=\frac{2 M_T}{v_2^2 \lambda_2}U^*\cdot{\rm diag}(m_1,m_2,m_3)\cdot U^{\dagger}.
\end{eqnarray}
The running of the soft-SUSY breaking slepton mass matrix 
$(m^2_{\widetilde{L}})_{ij}$ is proportional to the parameter 
combination $(Y_T^\dag Y_T)_{ij}$. \footnote{For the triplet-only
case. For the ${\bf 15}$ case we assume $Y_Z=Y_T$ at $M_G$, see the previous 
subsection.} This combination can again be expressed in terms of 
low-energy neutrino observables times an unknown scale:
\begin{eqnarray}\label{expressYuk}
(Y_T^\dag Y_T)_{ij}
&=&
\left[
\frac{2 M_T}{v_2^2 \lambda_2}
\right]^2
\left(
U\cdot{\rm diag}(m_1^2,m_2^2,m_3^2)\cdot U^{\dagger}
\right)_{ij}
\nonumber\\
&\equiv&
\widetilde{m}^{-2}
\sum_{k}U_{ik} U_{jk}^* m_k^2,
\end{eqnarray}
The different off-diagonal entries are explicitly given as
\begin{eqnarray}
(Y_T^\dag Y_T)_{12}
&=&
\widetilde{m}^{-2}
\left[
U_{11} U^*_{21} m_1^2+
U_{12} U^*_{22} m_2^2+
U_{13} U^*_{23} m_3^2
\right],\\ \nonumber
(Y_T^\dag Y_T)_{13}
&=&
\widetilde{m}^{-2}
\left[
U_{11} U^*_{31} m_1^2+
U_{12} U^*_{32} m_2^2+
U_{13} U^*_{33} m_3^2
\right],\\ \nonumber
(Y_T^\dag Y_T)_{23}
&=&
\widetilde{m}^{-2}
\left[
U_{21} U^*_{31} m_1^2+
U_{22} U^*_{32} m_2^2+
U_{23} U^*_{33} m_3^2
\right].
\end{eqnarray}
Inserting the convention for the matrix $U$ from eq. (\ref{def:unu}) 
results in 
\begin{eqnarray}\label{ydagy2}
(Y_T^\dag Y_T)_{12} &\propto& 
c_{12}s_{12}c_{13}c_{23}(m_2^2-m_1^2)-c_{13}s_{13}s_{23}e^{-i\delta}
\{(m_3^2-m_2^2) + c_{12}^2(m_2^2-m_1^2) \} ,
\\ \nonumber
(Y_T^\dag Y_T)_{13} &\propto& 
c_{12}s_{12}c_{13}s_{23}(m_1^2-m_2^2)-c_{13}s_{13}c_{23}e^{-i\delta}
\{(m_3^2-m_2^2) + c_{12}^2(m_2^2-m_1^2) \} , \\ \nonumber
(Y_T^\dag Y_T)_{23} &\propto& s_{23}c_{23}
\Big( (s_{12}^2-c_{12}^2)(m_2^2-m_1^2) + 
c_{13}^2\{(m_3^2-m_2^2)+c_{12}^2(m_2^2-m_1^2)\} \Big) \\ \nonumber
&-& s_{12}c_{12}s_{13}(c_{23}^2e^{-i\delta}-s_{23}^2e^{i\delta})(m_2^2-m_1^2). 
\end{eqnarray}
Note, that the off-diagonals can be expressed as a function of 
mass squared differences only, i.e. there is no dependence on the 
overall neutrino mass scale. However, again note that 
eq. (\ref{ydagy2}) depends on the energy scale, see the discussion 
below eq. (\ref{eq:ssII}) and in section (\ref{sec:num}). 
Also it is worth mentioning that with 
the convention of $U$ from eq. (\ref{def:unu}) the Majorana phases 
cancel in eq. (\ref{ydagy2}). 

As a starting approximation for the following discussion, let us 
assume that the lepton mixing matrix has exact tri-bimaximal (TBM)
form~\cite{Harrison:2002er}
\begin{equation}\label{tbm}
{U}={U}_{\textrm{TBM}}=
\begin{pmatrix}
\sqrt{\frac{2}{3}} & \frac{1}{\sqrt{3}} & 0 \\
-\frac{1}{\sqrt{6}} & \frac{1}{\sqrt{3}} & \frac{1}{\sqrt{2}} \\
 \frac{1}{\sqrt{6}} & -\frac{1}{\sqrt{3}} & \frac{1}{\sqrt{2}} 
\end{pmatrix} .
\end{equation}
As is well-known, eq.~(\ref{tbm}) is an excellent first-order 
approximation to the measured neutrino mixing angles \cite{Maltoni:2004ei}. 
For these values eq. (\ref{ydagy2}) simplifies to
\begin{eqnarray}\label{smplst}
|(Y_T^\dag Y_T)_{12}| &\propto&
\frac{1}{3} \Delta m_{\odot}^2 \\ \nonumber
|(Y_T^\dag Y_T)_{13}| &\propto&
\frac{1}{3}\Delta m_{\odot}^2,\\ \nonumber
|(Y_T^\dag Y_T)_{23}| &\propto&
\frac{1}{2} \Delta m_{\rm Atm}^2.
\end{eqnarray}
The ratios $(r^{12}_{13})=1$ and $(r^{12}_{23})=(r^{13}_{23})=\frac{2}{3}
\frac{\Delta m_{\odot}^2}{\Delta m_{\rm Atm}^2}$ result.

\begin{figure}[htb] \centering
\includegraphics[height=50mm,width=80mm]{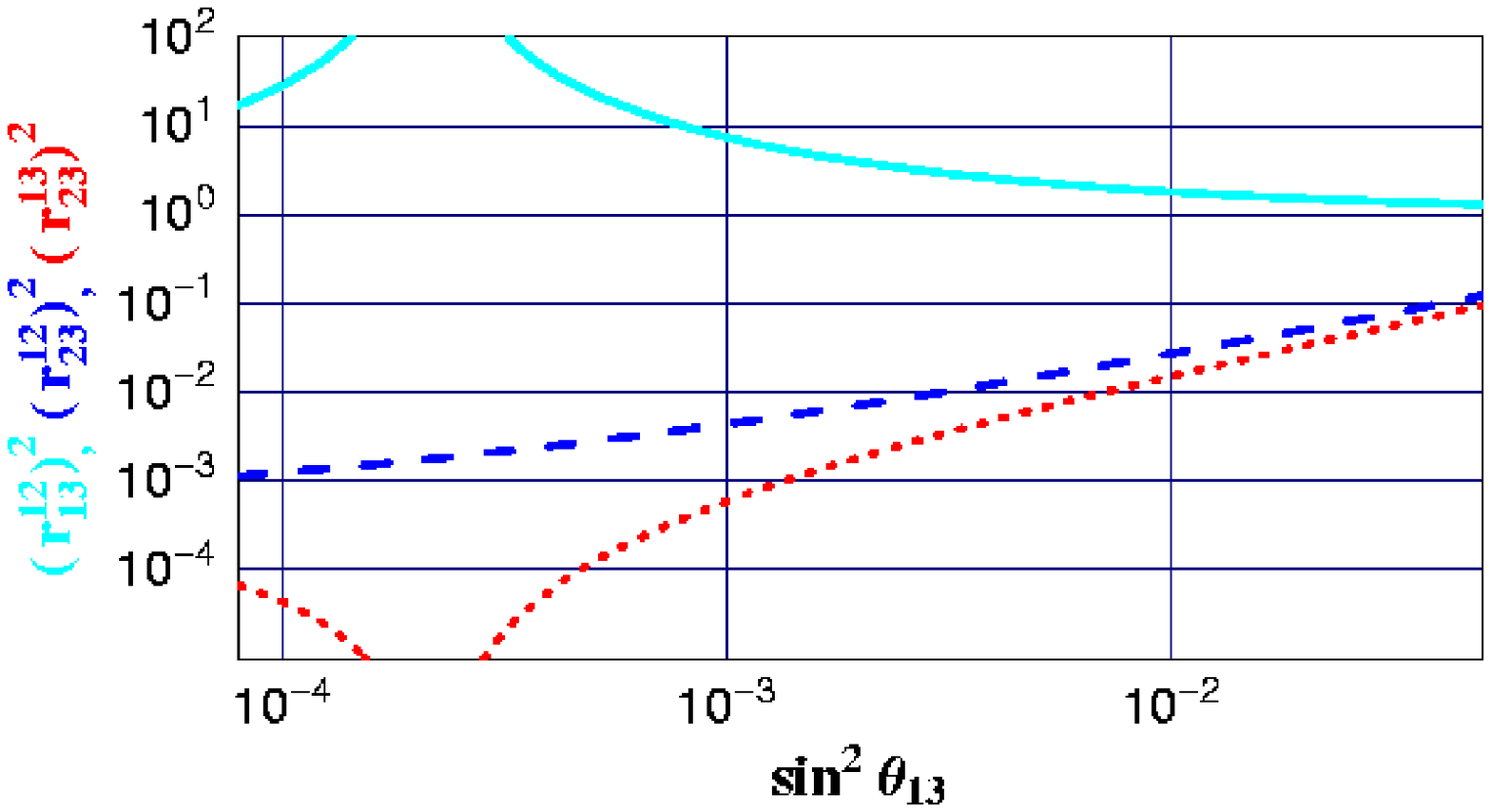} 
\includegraphics[height=50mm,width=80mm]{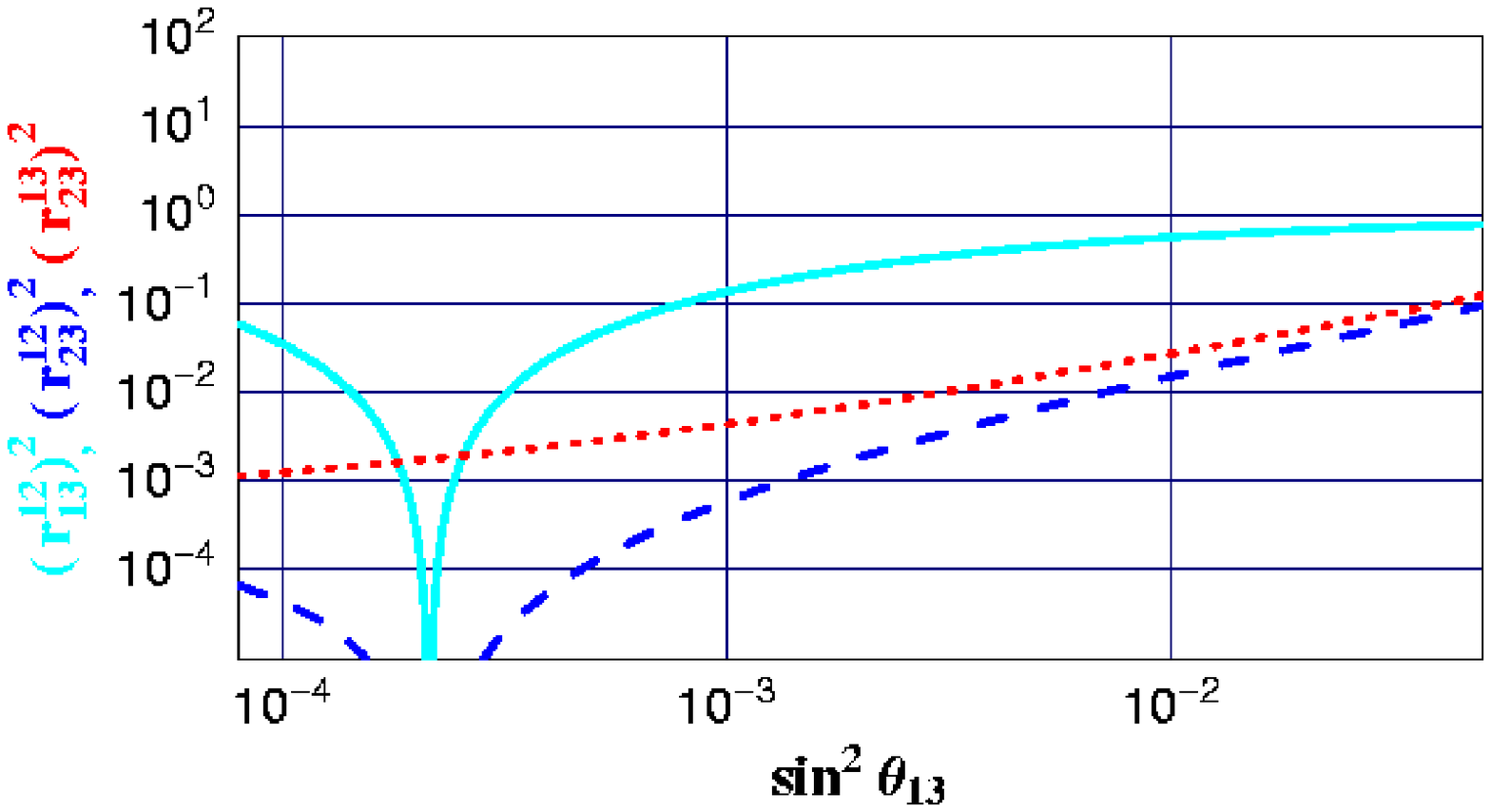}\\
\includegraphics[height=50mm,width=80mm]{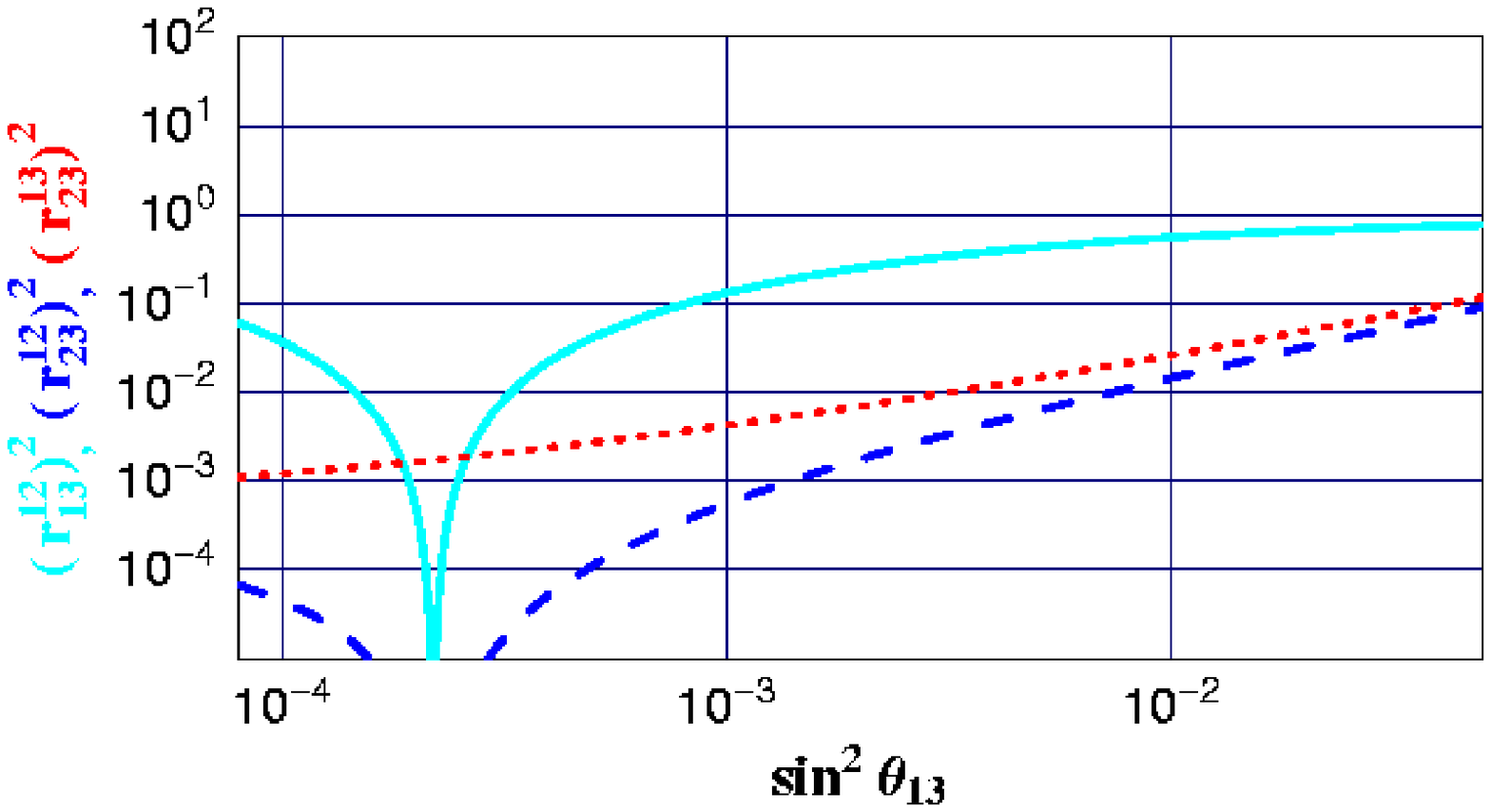} 
\includegraphics[height=50mm,width=80mm]{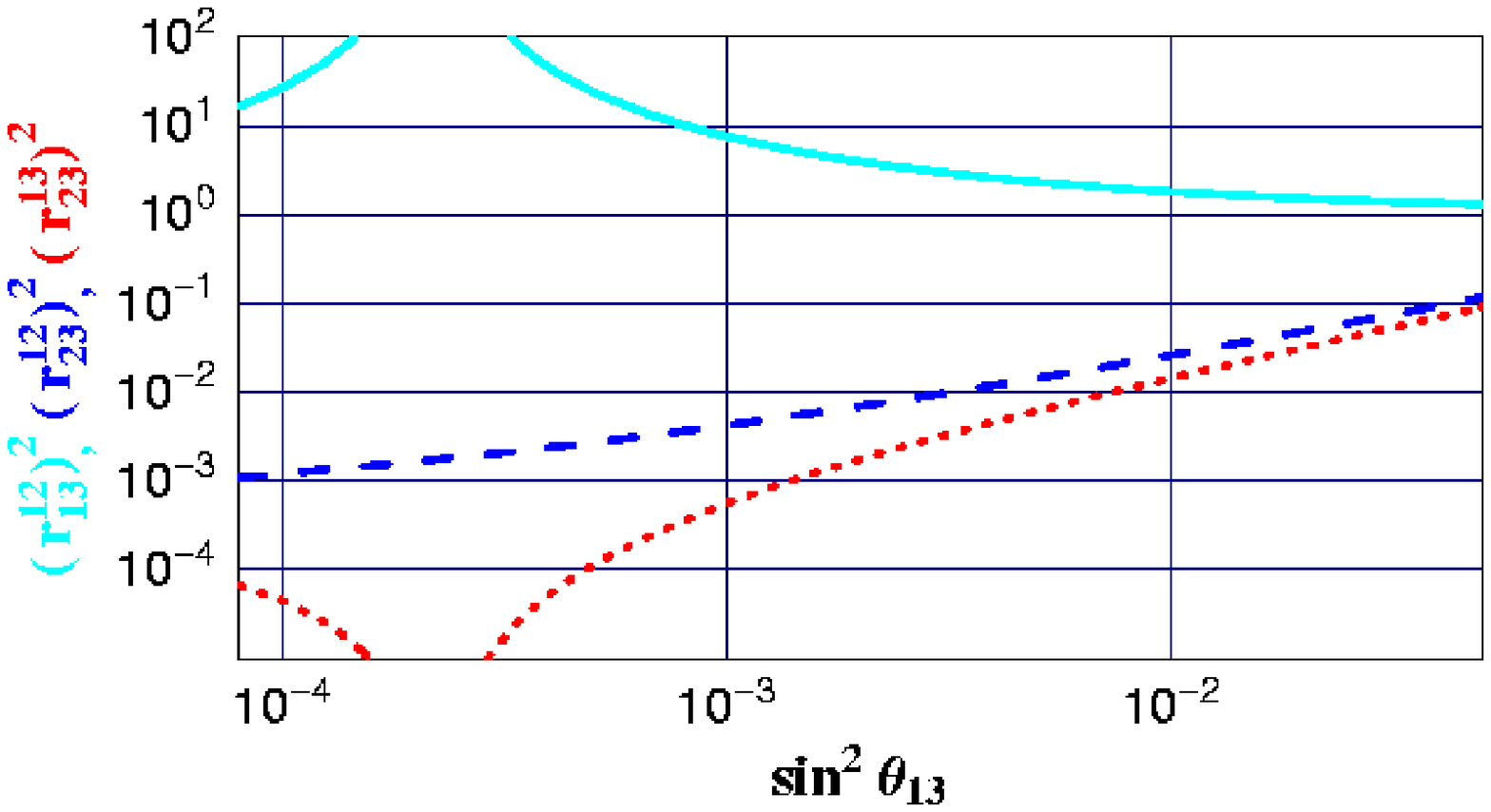}\\
\vskip-5mm
\caption{\label{fig:r-vs-s13-degM}Square ratios $(r^{12}_{13})^2$ (light 
blue line, full line), $(r^{12}_{23})^2$ (blue line, dashed line) and 
$(r^{13}_{23})^2$ (red line, dotted line) versus $s_{13}^2$ for NH 
(upper panels), IH (lower panels) for $\delta=0$ (left panels) and 
$\delta=\pi$ (right panels). The other light neutrino parameters have 
been fixed to their b.f.p. values. Note, that for $\tan^2\theta_{\rm A}=1$, 
$(r^{12}_{23})^2$ and $(r^{13}_{23})^2$ are symmetric under the exchange 
of $\delta=0$ $\leftrightarrow$ $\delta=\pi$. Also the simultaneous 
exchange of NH $\leftrightarrow$ IH and $(\delta=0$) $\leftrightarrow$ 
($\delta=\pi$) leads to the same values for the different $r^{ij}_{kl}$, 
in case of $\tan^2\theta_{\rm A}=1$.}
\end{figure}

For the general mixing matrix one can derive $(r^{ij}_{kl})^2$ 
using eq. (\ref{ydagy2}). For the currently allowed ranges of the 
neutrino parameters, the most important unknown turns out to be $s_{13}$, 
as fig. (\ref{fig:r-vs-s13-degM}) demonstrates.  In this figure 
$(r^{ij}_{kl})^2$ are shown as function of $s_{13}^2$ for 
$\tan^2\theta_{\rm A}=1$ and $\tan^2\theta_{\odot}=1/2$, as well as 
for the $\Delta m^2$ fixed at their best fit point values 
\cite{Maltoni:2004ei}. Currently $s_{13}^2\le 0.05$ at 3 $\sigma$ c.l. 
$r^{ij}_{kl}$ strongly depend on the value of $s_{13}$ and there exists 
a special value of $s_{13}$, for which either $(r^{12}_{23})$ or 
$(r^{13}_{23})$ even vanish, due to a cancellation between the different 
terms in eq. (\ref{ydagy2}). Note, however, that $(r^{12}_{23})$ and 
$(r^{13}_{23})$ can not vanish simultaneously. 
Note also, that for $\tan^2\theta_{\rm A}=1$, $(r^{12}_{23})^2$ and 
$(r^{13}_{23})^2$ are symmetric under the exchange of $\delta=0$ 
$\leftrightarrow$ $\delta=\pi$. Also, for non-zero values of $s_{13}$ 
the results depend on the assumed hierarchy of the left neutrinos and 
the simultaneous exchange of the cases (normal hierarchy) NH 
$\leftrightarrow$ IH (inverse hierarchy) and $(\delta=0$) 
$\leftrightarrow$ ($\delta=\pi$) leads to the same values for the 
different $r^{ij}_{kl}$ in case of $\tan^2\theta_{\rm A}=1$.

\begin{table}[htdp]
\begin{center}
\begin{tabular}{|c|c||c|c||c|c|}
\hline
 \multicolumn{2}{|c||}{} 
& \multicolumn{2}{|c||}{NH} 
&  \multicolumn{2}{|c|}{IH}
\\ \cline{3-6}
 \multicolumn{2}{|c||}{} 
& $\delta=0$ & $\delta=\pi$
& $\delta=0$ & $\delta=\pi$
\\ \hline \hline
$s_{12}=1/\sqrt{3}$ & $(r^{12}_{13})^2$ &
$1$ & 
& 
$1$ & 
$$ 
\\ \cline{2-6}
$s_{23}=1/\sqrt{2}$ & 
$(r^{12}_{23})^2$ &
$[2.8,7.4]\times 10^{-4}$ & 
$$ & 
$[2.8,7.2]\times 10^{-4}$ & 
$$
 \\ \cline{2-6} 
$s_{13}=0$ & 
$(r^{13}_{23})^2$ &
$[2.8,7.4]\times 10^{-4}$ & 
$$ & 
$[2.8,7.2]\times 10^{-4}$ & 
$$
 \\ \hline \hline
$s_{12}= 1/\sqrt{3}$ & $(r^{12}_{13})^2$ &
$[1.2,1.4]$ & 
$[0.71,0.81]$ & 
$[0.71,0.81]$ & 
$[1.2,1.4]$ 
\\  \cline{2-6}
$s_{23}=1/\sqrt{2}$ & $(r^{12}_{23})^2$ & 
$[0.12,0.13]$ & 
$[0.091,0.096]$ & 
$[0.085,0.093]$ & 
$[0.11,0.12]$
 \\  \cline{2-6}
$s_{13}=s_{13}^{\rm max}$ & $(r^{13}_{23})^2$ & 
$[0.091,0.096]$ & 
$[0.12,0.13]$ & 
$[0.11,0.12]$ & 
$[0.085,0.093]$
 \\ \hline \hline
 $s_{12}\neq 1/\sqrt{3}$ & $(r^{12}_{13})^2$ &
$[0.49,1.94]$ & 
$$ & 
$[0.49,1.94]$ & 
$$ 
\\ \cline{2-6}
$s_{23}\neq 1/\sqrt{2}$ & 
$(r^{12}_{23})^2$ &
$[1.8,12]\times 10^{-4}$ & 
$$ & 
$[1.8,12]\times 10^{-4}$ & 
$$
 \\ \cline{2-6} 
$s_{13}=0$ & 
$(r^{13}_{23})^2$ &
$[1.8,12]\times 10^{-4}$ & 
$$ & 
$[1.8,12]\times 10^{-4}$ & 
$$
 \\ \hline \hline
$s_{12}\neq 1/\sqrt{3}$ & $(r^{12}_{13})^2$ &
$[0.63,3.0]$ & 
$[0.35,1.7]$ & 
$[0.35,1.7]$ & 
$[0.63,3.0]$ 
\\  \cline{2-6}
$s_{23}\neq 1/\sqrt{2}$ & $(r^{12}_{23})^2$ & 
$[0.094,0.18]$ & 
$[0.062,0.15]$ & 
$[0.060,0.15]$ & 
$[0.089,0.18]$
 \\  \cline{2-6}
$s_{13}=s_{13}^{\rm max}$ & $(r^{13}_{23})^2$ & 
$[0.061,0.15]$ & 
$[0.093,0.18]$ & 
$[0.088,0.17]$ & 
$[0.058,0.14]$
 \\ \hline

\end{tabular}
\end{center}
\caption{
The parameters $r^{ij}_{kl}$ are given for several values of the neutrino 
mixing angles. $s_{13}^{\rm max}$ is the experimentally allowed maximum 
value: $(s_{13}^{\rm max})^2=0.050$ at $(3\sigma)$ c.l. NH and IH are 
normal and inverted hierarchy of neutrino masses, respectively. The 
intervals correspond to $(3\sigma)$ experimental allowed range of 
neutrino oscillation parameters: $s_{12}^2=0.26-0.40$, $s_{23}^2=0.34-0.67$, 
$\Delta m_{\odot}^2=(7.1-8.3)\times 10^{-5} {\rm eV^2}$ and 
$\Delta m_{\rm Atm}^2=(2.0-2.8)\times 10^{-3} {\rm eV^2}$. In the top two 
rows only the mass squared splittings are varied, while for the lower 
set also angles are allowed to vary.}
\label{tab:1}
\end{table}

Table (\ref{tab:1}) shows the currently allowed ranges for the 
$(r^{ij}_{kl})^2$ for $s_{13}=0$ and $s_{13}^{\rm max}$ for different 
assumptions about the remaining neutrino parameters for the different 
cases of NH and IH. These values serve to indicate the allowed 
variations for $r^{ij}_{kl}$ due to other parameters than $s_{13}$. 
As stated above, the allowed variation on $s_{13}$ is most important 
for the ``uncertainties'' in $(r^{ij}_{kl})^2$. However, also the 
current error bar on $\tan^2\theta_{\rm A}$ leads to a sizeable variation 
on $r^{ij}_{kl}$. Since $\Delta m^2_{\rm Atm}$ and $\Delta m^2_{\odot}$ 
are now known to much better precision than the neutrino angles, 
their variation is much less important for the $(r^{ij}_{kl})^2$, as 
table (\ref{tab:1}) demonstrates.

Finally, recall that all results presented in this subsection are 
based on the assumption that one of the extreme cases, $Y_Z=Y_T$ or 
$Y_Z=0$, is realized. The former corresponds to the SU(5) inspired 
model with a complete ${\bf 15}$ of section (\ref{sec:su5}), whereas the 
latter corresponds to the simplest triplet-only model discussed in 
section (\ref{sec:triplet}). However, we stress that departures in the 
ratios of LFV branching ratios from the values calculated in this 
subsection should not be interpreted as ``ruling out'' the 
seesaw type-II. Rather they should be interpreted in the sense that 
one has to go beyond minimal scenarios.

\section{Numerical results}
\label{sec:num}

In this section we present our numerical calculations. All results 
presented below have been obtained with the lepton flavour violating 
version of the program package SPheno \cite{Porod:2003um}. Calculations 
are done for the 15-plet case, using the assumption $Y_Z=Y_T$ at
$M_G$, as discussed above. Unless mentioned otherwise, we fit neutrino 
mass squared differences to their best fit values \cite{Maltoni:2004ei} 
and the angle to TBM values. Our numerical procedure is as 
follows. Inverting the seesaw equation, see eq.~(\ref{eq:ssII}), one 
can get a first guess of the Yukawa couplings for any fixed values of the 
light neutrino masses (and angles) as a function of the corresponding 
triplet mass for any fixed value of $\lambda_2$. This first 
guess will not give the correct Yukawa couplings, since the neutrino 
masses and mixing angles are measured at low energy, whereas for the 
calculation of $m_\nu$ we need to insert the parameters at the 
high energy scale. However, we can use this first guess to run 
numerically the RGEs to obtain the exact neutrino masses and angles 
(at low energies) for these input parameters. The difference between 
the results obtained numerically and the input numbers can then be 
minimized in a simple iterative procedure until convergence is achieved. 
As long as neutrino Yukawas are not too 
close to one we reach convergence in a few steps. However, in seesaw 
type-II the Yukawas run stronger than in seesaw type-I, thus our initial 
guess can deviate sizeably from the exact Yukawas. Since neutrino 
data requires at least one neutrino mass to be larger than about 
$0.05$ eV, we do not find any solutions for $M_T \gsim {\lambda_2} 
\cdot 10^{15}$ GeV. 

We have implemented the effects of the additional triplets (15-plets) 
including the two-loop contributions to the RGEs for gauge couplings 
and gaugino masses, one-loop contributions to the remaining MSSM 
parameters and one-loop RGES for the new parameters in SPheno. For 
consistency we have also included 1-loop threshold corrections for 
gauge couplings and gaugino mass parameters at the scale corresponding 
to the mass of the triplet. The MSSM part is implemented at the 
2-loop level and, thus, in principle one should also include the effect 
of the 15-plets consistently for all parameters at this level. 
However, the correct fit of the neutrino data require that either 
the triplet (15-plet) Yukawa couplings are small and/or that $M_T$ is 
close to $M_G$ implying that the ratio $M_T/M_G$ is significantly 
smaller than $M_G/m_Z$ and thus one expects only small effects.

\subsection{Numerical results for LFV}

The analytical results presented in the previous section allow to 
estimate ratios of branching ratios for LFV decays. For absolute 
values of the branching ratios, as well as for cross-checking the 
reliability of the analytical estimates, one has to resort to a 
numerical calculation. Below we show results only for a few 
``standard'' mSugra points, namely for SPS3 \cite{Allanach:2002nj} 
and SPS1a' \cite{AguilarSaavedra:2005pw}. However, we have checked 
with a number of other points that our results for ratios of branching 
ratios are generally valid. 

\begin{figure}[htb] \centering
\includegraphics[height=50mm,width=80mm]{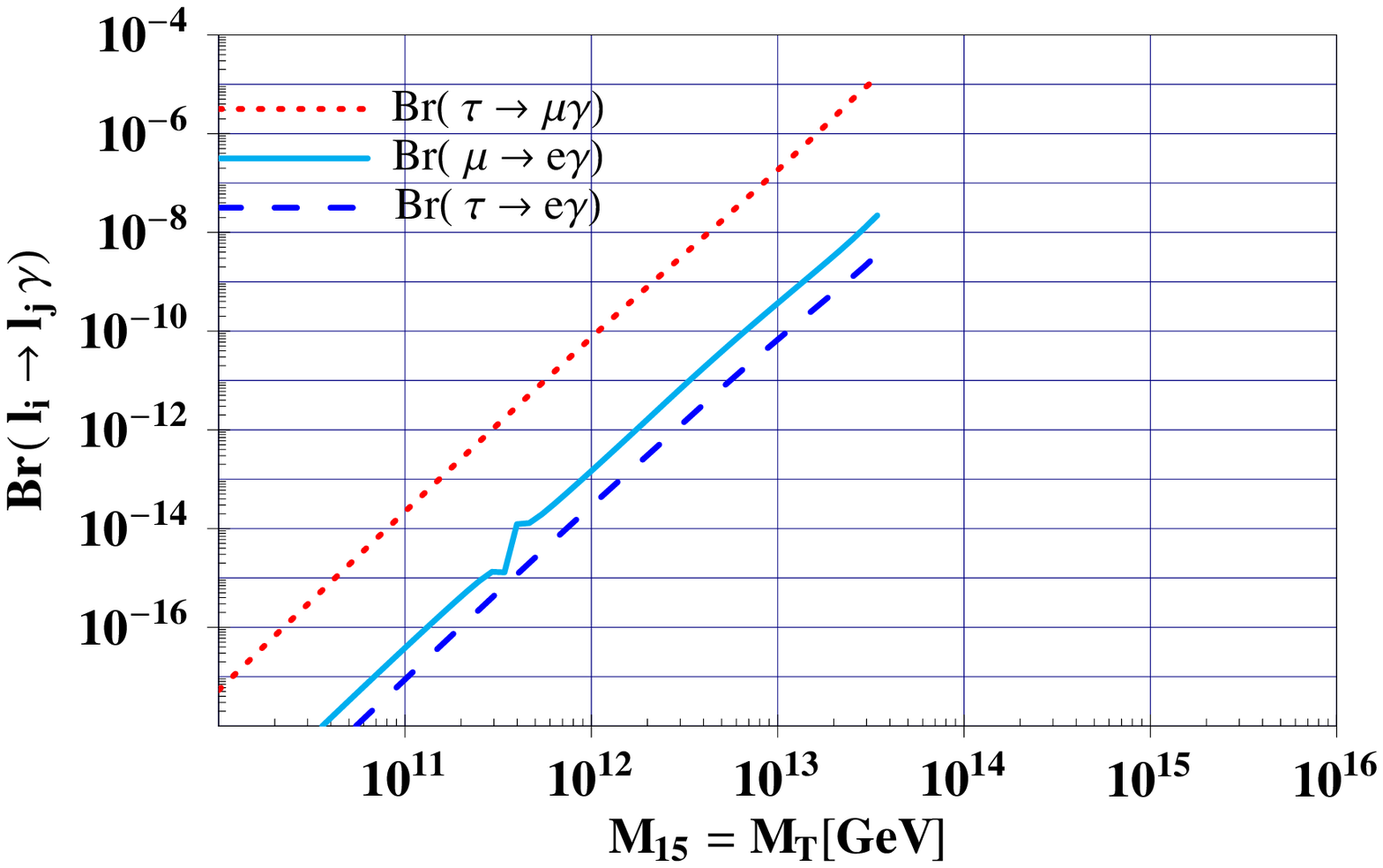}
\includegraphics[height=50mm,width=80mm]{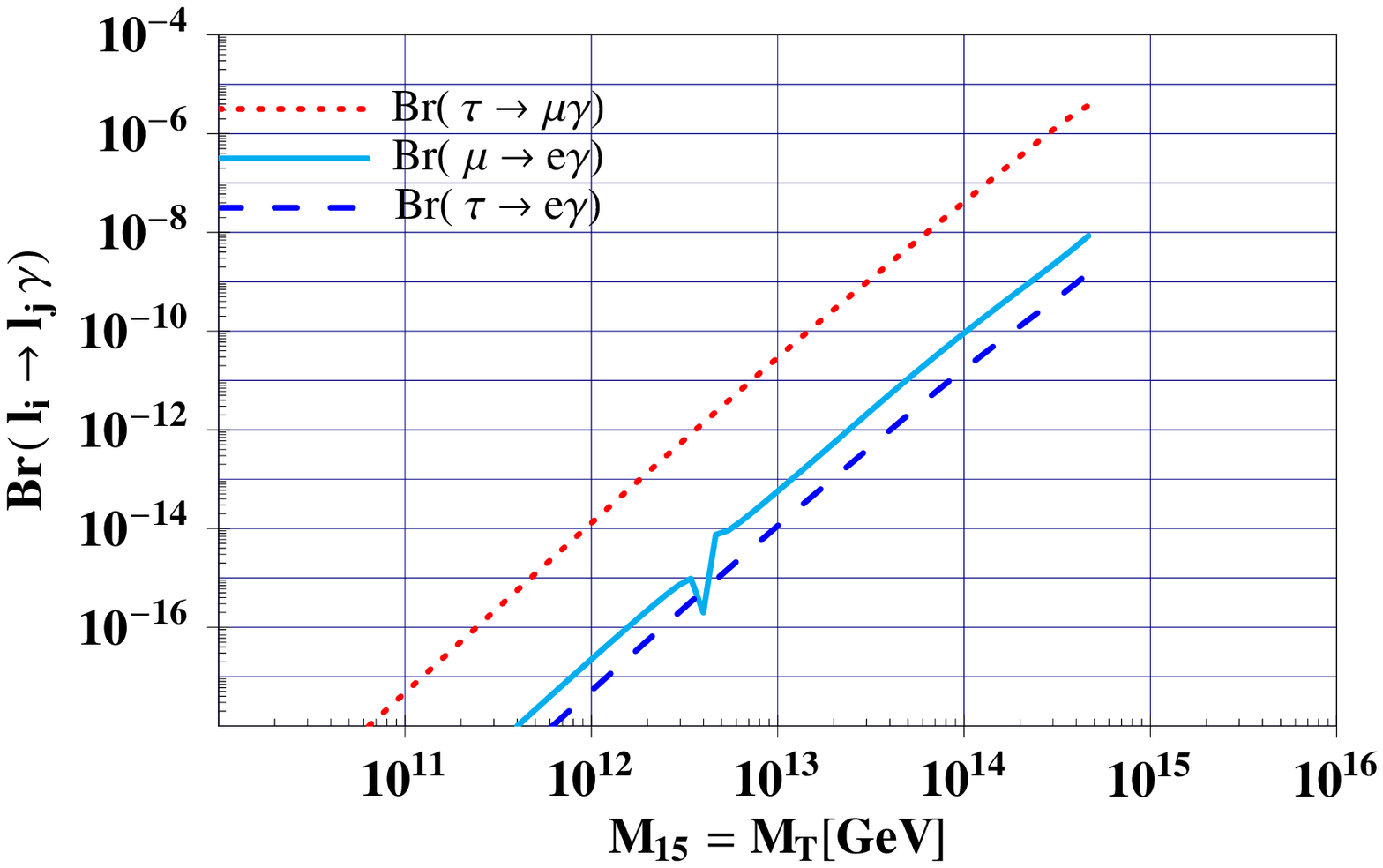} \\
\includegraphics[height=50mm,width=80mm]{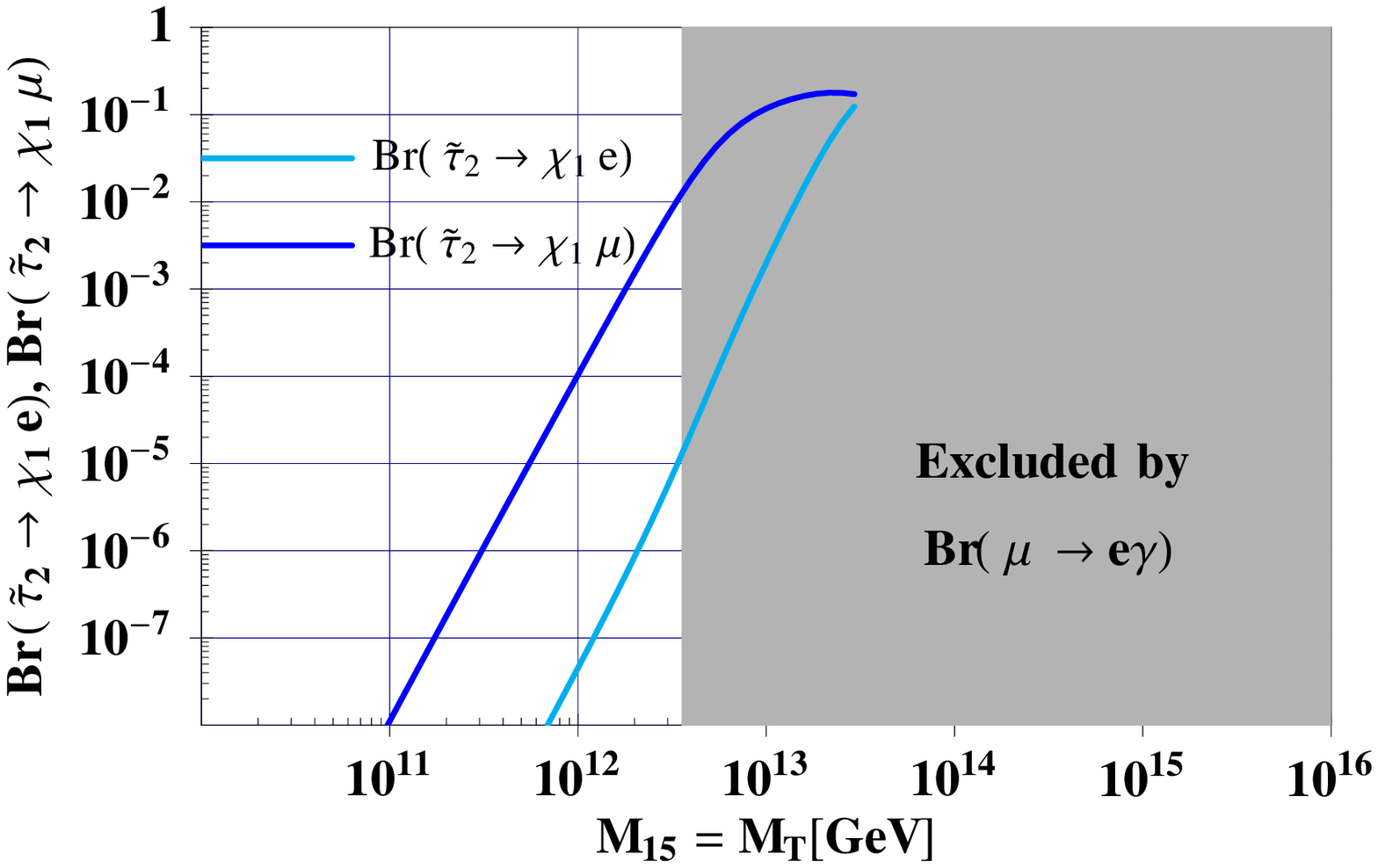}
\includegraphics[height=50mm,width=80mm]{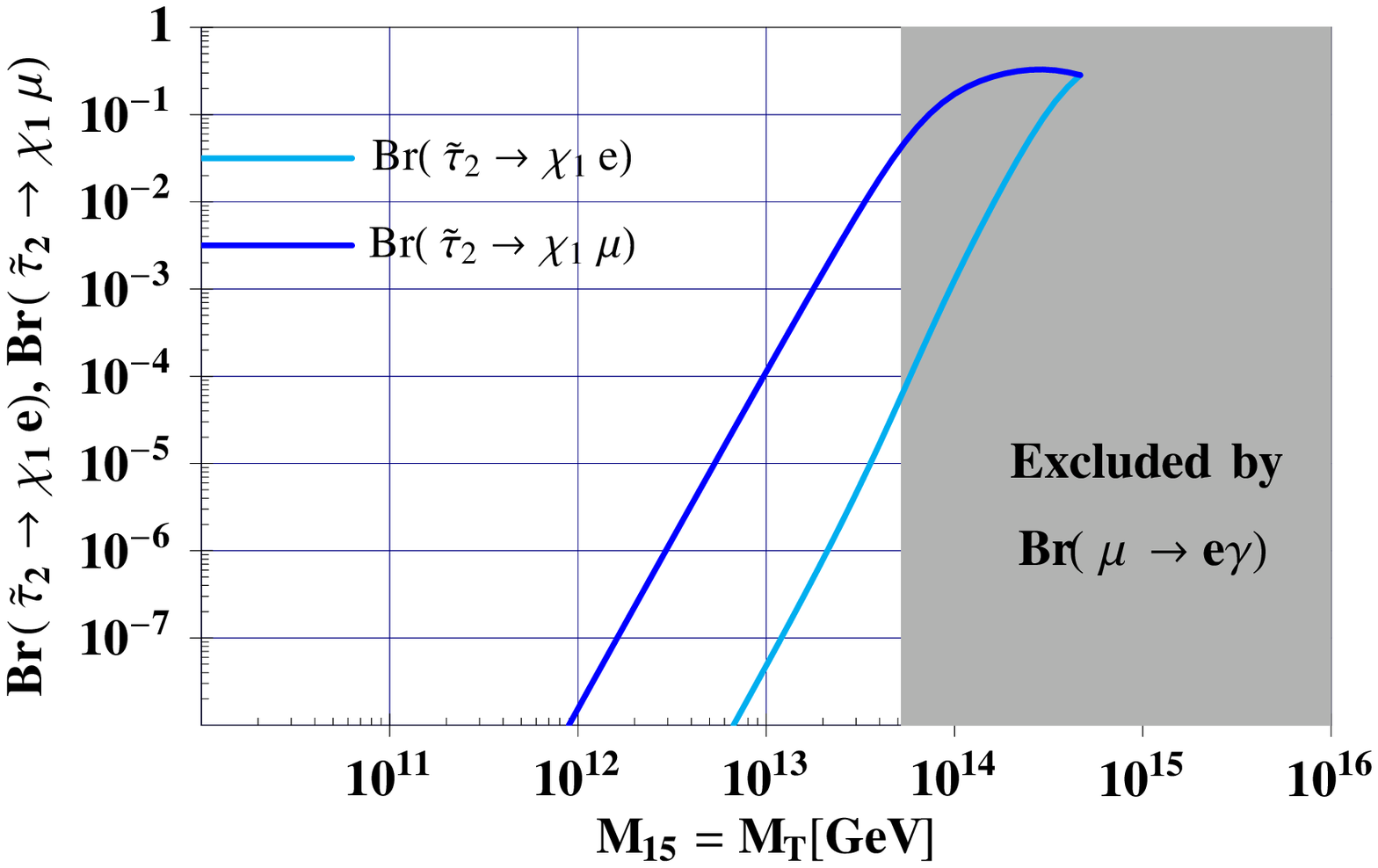} \\
\vskip-5mm
\caption{\label{fig:sps3lfv}Lepton flavour violating branching ratios versus 
$M_T=M_{15}$ for the standard mSugra point SPS3 for two values of 
$\lambda_2$. To the left $\lambda_2=0.05$, to the right $\lambda_2=0.5$. 
The plots show $Br(l_i\to l_j+\gamma)$ (top) and 
$Br({\tilde\tau_2}\to e,\mu +\chi_1^0)$ (bottom). 
Ratios of the different branching ratios 
follow closely the analytical expectations. The regios excluded by the 
current upper limit on $Br(\mu \to e+\gamma)$ is shown also in the lower 
plot.}
\end{figure}

Fig. (\ref{fig:sps3lfv}) shows examples of LFV decays for the mSugra 
point SPS3 as a function of $M_T=M_{15}$ for two different values of 
$\lambda_2$. The upper plots show $Br(l_i \to l_j+\gamma)$, while the 
lower ones show $Br({\tilde\tau_2}\to e,\mu +\chi_1^0)$. We 
have also calculated $Br(l_i \to 3 l_j)$, but these are not shown in 
the plots, because they follow very well the approximate 
relation \cite{Arganda:2005ji,Hisano:1998fj}
\begin{equation}\label{relateLFV}
\frac{{\rm Br}(l_i \to 3 l_j)}{{\rm Br}(l_i \to l_j + \gamma)} \simeq 
\frac{\alpha}{3\pi}\Big(\log(\frac{m_{l_i}^2}{m_{l_j}^2})-\frac{11}{4}\Big).
\end{equation}
All LFV branching ratios show a very strong dependence on the value 
of $M_T$ and due to the stronger running of parameters in the seesaw 
type-II case, compared to the seesaw type-I, the dependence on 
the seesaw scale is stronger than in seesaw-I \cite{Hirsch:2008dy}. 
See also the discussion in section (\ref{sec:ana}). 

For the calculation shown in fig.~(\ref{fig:sps3lfv}), we have fitted 
the neutrino angles to exact tri-bimaximal values. One sees that, 
as long as the different LFV branching ratios are small, ratios of 
branching ratios are constants, which follow very well the analytical 
expectations. Currently the most important phenomenological constraints 
comes from the upper limit on $Br(\mu\to e+\gamma)$, 
$Br(\mu\to e+\gamma)\le 1.2 \cdot 10^{-11}$ \cite{pdg}. Note that the 
``dip'' in $Br(\mu\to e +\gamma)$ is due to a level-crossing of selectron 
and smuon mass eigenstates. \footnote{In mSugra the left-selectron is 
usually slightly lighter than the left-smuon. In the mSugra plus seesaw 
case, for both type-I and type-II seesaw, the additional Yukawas change 
the running of the slepton masses. For large Yukawas (i.e. large $M_T$) 
fitting current neutrino data requires couplings such that the smuon mass 
runs faster to smaller values than the selectron mass. However, the 
splitting between seletron and smuon mass eigenstates is expected to be 
too small to be measurable in most parts of the parameter space, see 
the discussion in the next subsection.} For SPS3 one 
finds that this limit rules out $Br({\tilde\tau_2}\rightarrow \mu +\chi_1^0)$ 
larger than a few percent, the exact number depending on the unknown 
parameter $\lambda_2$. Fig. (\ref{fig:sps3lfv}) to the left (right) shows 
results for $\lambda_2=0.05$ ($\lambda_2=0.5$). Recall that neutrino 
physics fixes only $M_T/{\lambda_2}$. However, note also that the upper 
limit on $Br({\tilde\tau_2}\rightarrow \mu +\chi_1^0)$ depends only 
weakly on $\lambda_2$.

\begin{figure}[htb] \centering
\includegraphics[height=50mm,width=80mm]{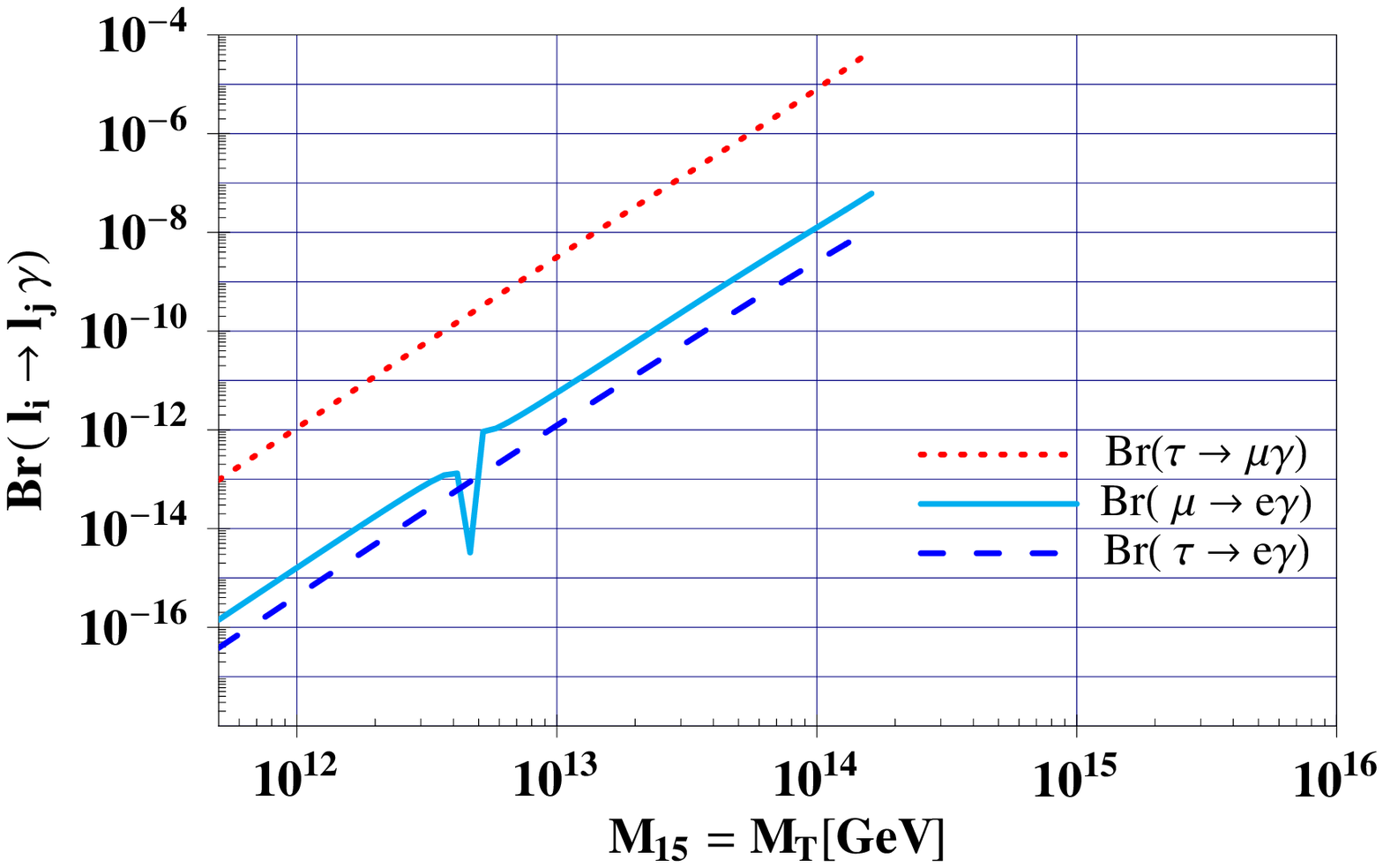} 
\includegraphics[height=50mm,width=80mm]{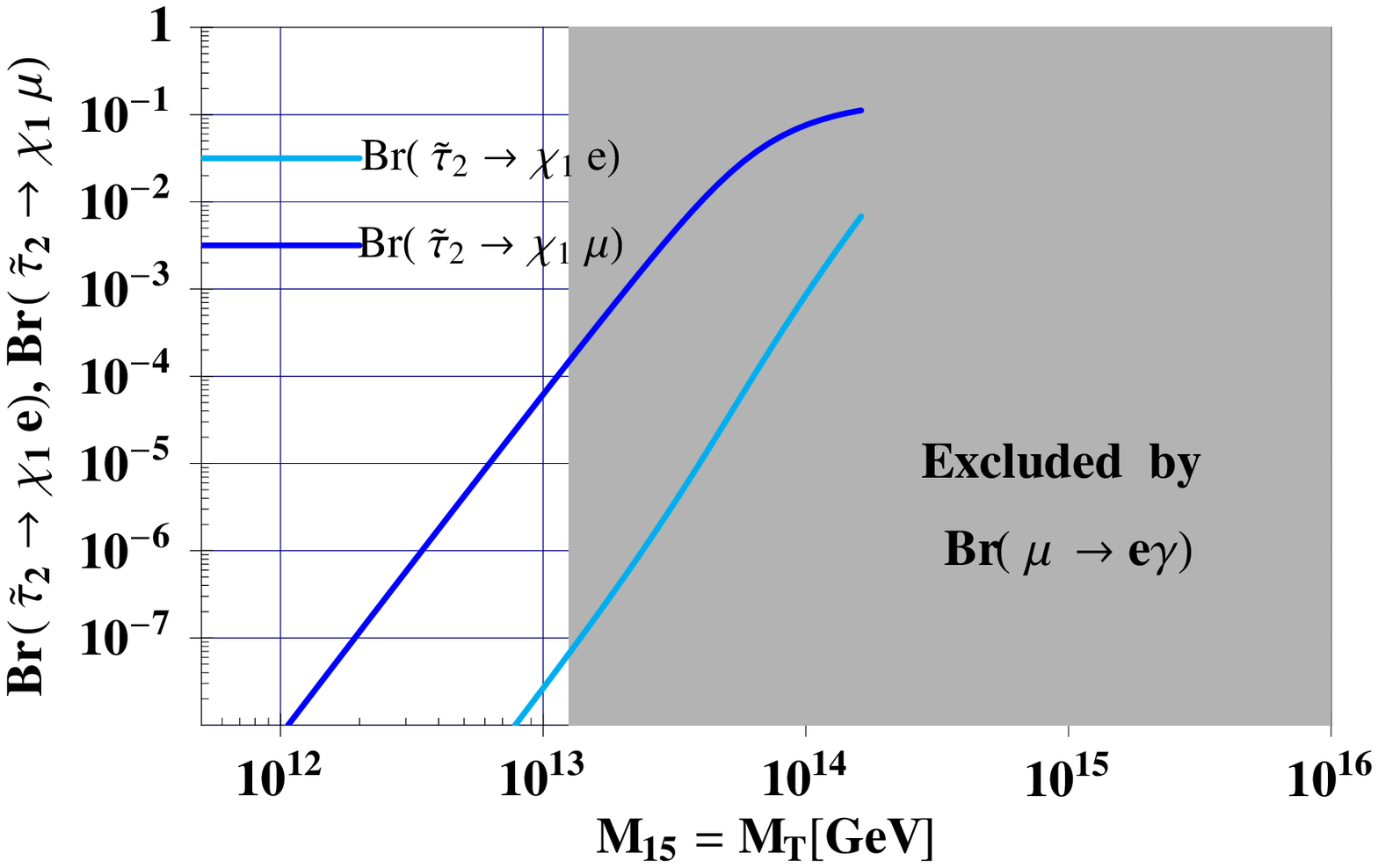} \\
\vskip-5mm
\caption{\label{fig:sps1lfv}As fig. (\ref{fig:sps3lfv}), but for the 
mSugra standard point SPS1a' and for $\lambda_2=0.5$. For a slepton 
spectrum as light as expected for SPS1a' $Br(\mu\to e+\gamma)$ rules 
out the possibility to observe large lepton flavour violating slepton 
decays at the LHC.}
\end{figure}

It is well-known that absolute values of LFV branching ratios depend 
very strongly on the SUSY spectrum, for example $Br(\mu \to e+\gamma)
\propto 1/m_{SUSY}^8$ \cite{Hisano:1995cp}. Since both left-sleptons as 
well as (lightest) neutralino and chargino are approximately a factor of 
two heavier for SPS3 than for SPS1a', one expects that $Br(\mu \to e+\gamma)$ 
gives a strong constraint on the observability of LFV at the LHC for 
SPS1a'. This is confirmed numerically, as shown in fig. (\ref{fig:sps1lfv}), 
which shows $Br(l_i \to l_j+\gamma)$ and $Br({\tilde\tau_2}\rightarrow e,\mu 
+\chi_1^0)$ as function of $M_T=M_{15}$ for the example of $\lambda_2=0.5$.
Given the current limit on $Br(\mu \to e+\gamma)$ one expects 
$Br({\tilde\tau_2}\rightarrow \mu +\chi_1^0)\lsim$ (few) $10^{-4}$. 
Note that again we have fitted neutrino angles to tri-bimaximal values 
in this calculation and that ratios of LVF branching ratios follow 
closely the analytical expressions.

\subsection{Sparticles Masses and seesaw scale}
\label{sec:scale}
As disdussed in the analytic section, the running of soft parameters 
allows, in principle, an indirect determination of the seesaw scale. 
In this section we discuss numerical results for the running of the 
``invariants'' defined above. Although below we show plots only 
for the combination $(m_{\tilde L}^2 -m_{\tilde E}^2)/M_1^2$ we 
have checked numerically that all invariants shown in 
fig. (\ref{fig:ana}) behave qualitatively in the same way.

\begin{figure}[htb] \centering
\includegraphics[height=50mm,width=80mm]{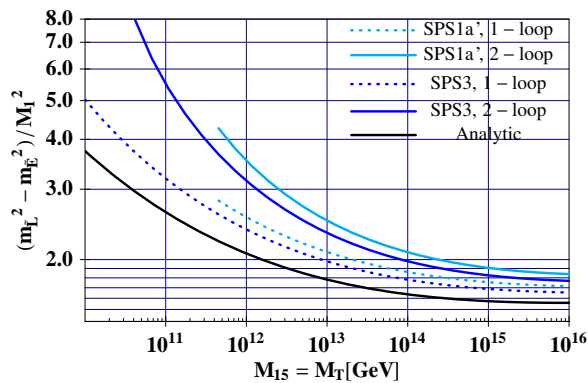}
\vskip-5mm
\caption{\label{fig:comp}''Invariant'' 
$(m_{\tilde L}^2 -m_{\tilde E}^2)/M_1^2$, calculated with negligibly 
small Yukawa couplings for two mSugra standard points. The figure 
shows a comparsion of different calculation. The curve labeled ``Analytic'' 
uses the formulas presented in the previous section. 1-loop and 2-loop 
stand for exactly solved numerical calculations using 1-loop and 
2-loop RGEs. Note the significant shift when going from 1-loop order 
to 2-loop order.}
\end{figure}

Fig. (\ref{fig:comp}) shows $(m_{\tilde L}^2 -m_{\tilde E}^2)/M_1^2$ 
as a function of $M_T=M_{15}$ for SPS1a' and SPS3 comparing different 
calculations. This plot assumes that the Yukawas of the 15-plet are 
negligibly small, i.e. neutrino mass are {\em not correctly fitted 
in this calculation}. The black line is the analytical calculation 
based on 1-loop RGEs and the leading-log approximation with an assumed 
$m_{SUSY}=1$ TeV. The dotted lines are the numerically exact results 
for this invariant using 1-loop RGEs, while the full lines are the 
exact results using 2-loop RGEs. Obviously the ``invariant'' 
does depend to a certain degree on the mSugra point, as already 
pointed out in section (\ref{sec:ana}). However, we also find a 
considerable upward shift of $(m_{\tilde L}^2 -m_{\tilde E}^2)/M_1^2$, 
when going from the 1-loop to the 2-loop calculation. Since the 
dependence of $(m_{\tilde L}^2 -m_{\tilde E}^2)/M_1^2$ on the value 
of $M_T$ is only logarithmic, even such a moderate change in the 
invariant is important, if one wants to extract an indirect estimate 
on $M_T$ from such a measurement. Note that for the point SPS1a' 
the calculation stops at $M_{15}\sim 10^{11.6}$ GeV, the lowest value 
of $M_{15}$ for which correct electro-weak symmetry breaking occurs.

We have checked by an exact numerical calculation that the other 
invariants shown in section (\ref{sec:ana}) suffer from similar 
changes when going from 1-loop order to 2-loop. In other words, 
if one wants to learn about the seesaw scale from measurements of 
the soft masses, a careful analysis at 2-loop order will be 
necessary. Also note that, due to the logarithmic dependence on 
$M_T$, highly precise measurements will be necessary, especially 
if $M_T$ is large, say $M_T \ge 10^{12-13}$ GeV.

\begin{figure}[htb] \centering
\includegraphics[height=50mm,width=80mm]{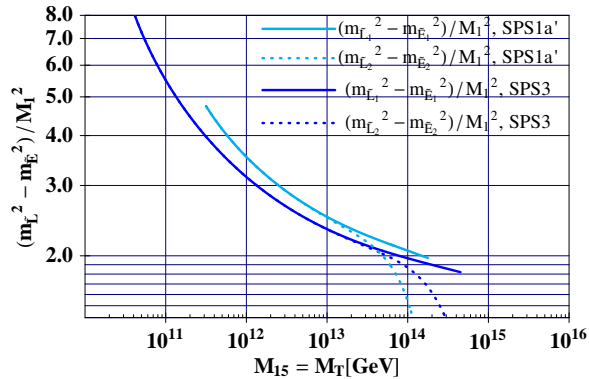}
\vskip-5mm
\caption{\label{fig:compYuk}''Invariant'' 
$(m_{\tilde L}^2 -m_{\tilde E}^2)/M_1^2$ calculated with Yukawa 
couplings fitted to neutrino data, for an arbitrary choice of 
$\lambda_2 = 0.5$. The calculation uses 2-loop RGEs. Results 
are shown for SPS1a' and SPS3. Neutrino angles are assumed to 
have exact TBM values.}
\end{figure}

Fig. (\ref{fig:compYuk}) shows $(m_{\tilde L}^2 -m_{\tilde E}^2)/M_1^2$ 
calculated with Yukawa couplings fitted to neutrino data, for an arbitrary 
choice of $\lambda_2 = 0.5$. The calculation uses 2-loop RGEs and 
results are shown again for the mSugra standard points SPS1a' and SPS3. 
For $M_T$ low, say $M_T \le 10^{13}$ GeV or so in this example, Yukawa 
couplings which explain current neutrino data are too small to induce 
any significant effect in the determination of 
$(m_{\tilde L}^2 -m_{\tilde E}^2)/M_1^2$. 

However, for larger values of $M_T$ sizeable differences between fig. 
(\ref{fig:comp}) and fig. (\ref{fig:compYuk}) show up. First of all, 
for negligibly small Yukawas the calculation can vary $M_T$ freely up 
to the GUT scale. If instead we insist to fit neutrino masses, such 
large values for $M_T$ are not allowed. The downward turn in 
$(m_{\tilde L}^2 -m_{\tilde E}^2)/M_1^2$ is due to Yukawas, which if 
larger than ${\cal O}(0.1)$ contribute sizeable in the running of 
the soft parameters. In the example shown in this figure 
$\lambda_2 = 0.5$ has been chosen. For smaller values of $\lambda_2$ 
again for fixed values of the Yukawa couplings to fit neutrino masses 
a lower $M_T$ is required. Correspondingly, for smaller $\lambda_2$ 
the effect of the Yukawas is seen for smaller values of $M_T$. 

It is also found that slepton mass parameters of the first and 
second generation run differently for large values of $M_T$, see 
fig. (\ref{fig:compYuk}). This difference can be traced to the fact 
that we have fitted neutrino angles to take exact TBM values. In 
this limit, $m_{\tilde L_1}^2 \propto \Delta m^2_{\odot}$, while 
$m_{\tilde L_2}^2 \propto \Delta m^2_{\rm Atm}$. Thus, at the largest 
values of $M_T$ sizeable mass differences between 1st and 2nd generation 
sleptons show up. This difference is expected to be smaller for 
non-zero values of $s_{13}$. Note, however, that for the example 
points shown in fig. (\ref{fig:compYuk}), there is the upper limit 
on $M_T$ from $Br(\mu\to e+\gamma)$, discussed in the last subsection. 
For SPS1a' $M_T\lsim 1.5\cdot 10^{13}$ GeV, for SPS3 $M_T\lsim 6\cdot 10^{13}$ 
GeV. This limits the range of $M_T$ where differences between 1st 
and 2nd generation slepton masses might be observable. We mention  
that a recent paper \cite{Allanach:2008ib} claims that mass differences 
between smuons and selectrons can be measured very accurately, even 
at the LHC. Depending on the mSugra point $(m^2_{\tilde \mu} -
 m^2_{\tilde e}) / (m^2_{\tilde \mu} + m^2_{\tilde e})$ 
as small as $O(10^{-4})$  might 
be measurable \cite{Allanach:2008ib} provided the leptons have sufficient
energy to pass the experimental cuts.

\begin{figure}[htb] \centering
\includegraphics[height=50mm,width=80mm]{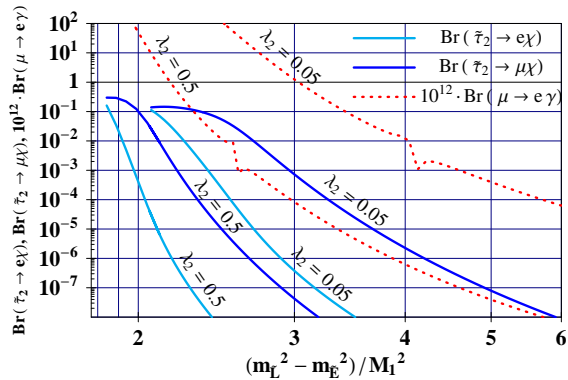}
\vskip-5mm
\caption{\label{fig:lfvvinv}Branching ratios for LFV decays versus 
$(m_{\tilde L}^2 -m_{\tilde E}^2)/M_1^2$ for SPS3 for two different 
values of $\lambda_2$. Measuring both types of observables allow 
in principle to disentangle $\lambda_2$ and $M_T$.}
\end{figure}

All observables discussed so far are sensitive only to a 
combination of $M_T$ and $\lambda_2$. If, however, both LFV 
decays as well as $(m_{\tilde L}^2 -m_{\tilde E}^2)/M_1^2$ 
could be measured in the future, one could disentangle the 
two parameters, in principle, by combining both measurements. 
This is demonstrated in fig. (\ref{fig:lfvvinv}), which shows LFV 
decays, $Br(\mu\to e +\gamma)$ and $Br({\tilde\tau_2}\rightarrow e,\mu 
+\chi_1^0)$ versus $(m_{\tilde L}^2 -m_{\tilde E}^2)/M_1^2$, 
for two different values of $\lambda_2$. Note again that the ``dip'' 
in $Br(\mu\to e +\gamma)$ is due to a level-crossing of selectron 
and smuon mass eigenstates. However, again we warn 
that a full 2-loop calculation is needed, before any quantitative 
conclusions could be drawn from such a measurement.

\section{Conclusions}
\label{sec:cncl}

We have studied phenomenological implications of the supersymmetric 
version of the type-II seesaw within mSugra. We have calculated lepton 
flavour violating observables, such as $Br(l_i\to l_j+\gamma)$ and LFV 
scalar tau decays. We have found branching ratios for LFV violating 
stau decays are large enough to be detectable at the LHC in principle. 
We have pointed out that in the simplest case of only one triplet 
coupling to the SM leptons, ratios of LFV branching ratios can be 
calculated from low energy neutrino data only. However, for the case 
of a complete ${\bf 15}$ multiplet the situation is not as straightforward. 
In the SU(5) inspired model the Yukawa couplings $Y_T$ and $Y_Z$ are 
related to the ${\bf Y_{15}}$ and the conclusions remain unchanged. 
However, allowing $Y_T$ and $Y_Z$ to be free parameters, the relation 
with neutrino physics is lost. Thus, seesaw type-II can not 
be ruled out by any LFV measurements in general. Instead measuring 
ratios of LFV branching ratios can be understood as a consistency 
check for the minimal seesaw type-II models.

We have also calculated the soft masses as a function of the 
seesaw parameters. As discussed in some detail, there are certain 
combinations of soft masses, which are approximately constants 
over large regions of mSugra space. These ``invariants'' contain 
indirect information about the seesaw scale. Measuring SUSY 
masses as precisely as possible will therefore allow to constrain 
the scale of seesaw type-II indirectly. However, theoretically 
there are many possibilities, why any single of the ``invariants'' 
we discussed could  depart from the simplest mSugra expectations. 
Only a consistent departure of several ``invariants'', together 
with measurements of LFV processes, could therefore be taken as 
a hint for seesaw type-II.

\section*{Acknowledgments}

Work supported by Spanish grants FPA2005-01269 and Accion Integrada
HA-2007-0090 (MEC) and by the European Commission network
MRTN-CT-2004-503369 and ILIAS/N6 RII3-CT-2004-506222.
 W.P.~is supported by the DAAD, project number D/07/13468,
 and partially by the German Ministry of Education and Research (BMBF)
under contract 05HT6WWA.

\begin{appendix}
\section{Contributions to the $\beta$ functions}

Using general formulas by \cite{Martin:1993zk} we obtain 
for the RGEs of the gauge couplings: 
\begin{eqnarray}
\frac{d g_a}{d t} &=& \frac{1}{16 \pi^2} B_a^{(1)} g^3_a 
+ \left( \frac{1}{16 \pi^2}\right)^2 g^3_a  \left( B_{ab}^{(2)} g^2_b +
C_{a}^{b} Tr\left( Y_b Y^\dagger_b\right) + D^{b}_{a} |\lambda_b|^2 \right) 
\end{eqnarray}
with
\begin{eqnarray}
B_1 &=& b_1 + \frac{3}{5}\left( \frac{8}{3} n_S + 3 n_T + \frac{1}{6} n_Z\right)  
\nonumber \\
B_2 &=& b_2 + \frac{3}{5}\left( \frac{8}{3} n_S + 3 n_T + \frac{1}{6} n_Z\right)  
\nonumber \\
B_3 &=& b_3 + \frac{3}{5}\left( \frac{8}{3} n_S + 3 n_T + \frac{1}{6} n_Z\right)  
 \\
B_{ab}^{(2)} &=& b_{ab}^{(2)} + b_{ab}^{(2,S)} n_S
              + b_{ab}^{(2,T)} n_T + b_{ab}^{(2,Z)} n_Z \\[2mm]
C_{a}^{u,d,l,S,T,Z} &=&\left( \begin{array}{cccccc}
\frac{26}{5} & \frac{14}{5} & \frac{18}{5} &  \frac{18}{5} &  \frac{27}{5}& \frac{14}{5}\\
6 & 6 & 2 & 0 & 7 & 6\\
4 & 4 & 0 & 6 & 0 & 4\\
\end{array}\right) \,\,\,,\,\,\,
D_a^{b} = \left( \begin{array}{cc}
\frac{27}{5} &\frac{27}{5} \\
7 & 7 \\ 0 & 0 \end{array}\right)
\end{eqnarray}
and $(b_1,b_2,b_3) = (33/5, 1, -3)$,
\begin{eqnarray}
b_{ab}^{(2)} = \left( \begin{array}{ccc}
\frac{199}{25} & \frac{27}{5} & \frac{88}{5} \\
\frac{9}{5} & 25 & 24 \\
\frac{11}{5} & 9 & 14 \\
\end{array}\right) \,\,,\,\,
b_{ab}^{(2,S)} = \left( \begin{array}{ccc}
\frac{128}{75} & 0 & \frac{64}{3} \\
0 & 0 & 0 \\
\frac{8}{3} & 0 & \frac{145}{3}
\end{array}\right)  \,\,,\,\,
b_{ab}^{(2,T)} = \left( \begin{array}{ccc}
\frac{108}{25} & \frac{72}{5} & 0 \\
\frac{24}{5} & 24 & 0 \\
0 & 0 & 0 
\end{array}\right)  \,\,,\,\,
b_{ab}^{(2,Z)} = \left( \begin{array}{ccc}
\frac{1}{150} & \frac{3}{10} & \frac{8}{15} \\
\frac{1}{10} & \frac{21}{2} & 8 \\
\frac{1}{15} & 3 & \frac{34}{3}
\end{array}\right) \nonumber \\
\end{eqnarray}
\end{appendix}

\end{document}